\documentclass[useAMS,usenatbib]{mn2e}
\usepackage[T1]{fontenc}
\usepackage{aecompl}
\usepackage{subcaption}
\usepackage[fleqn]{amsmath}
\usepackage{amssymb}
\usepackage{amsmath}
\usepackage{graphicx}
\usepackage{hyperref}
\usepackage{xcolor}
\hypersetup{
    colorlinks,
    linkcolor={red!70!black},
    citecolor={blue!70!black},
    urlcolor={blue!90!black}
}
\newcommand{\diff}{\mathrm{d}}
\newcommand{\prob}{\mathrm{Pr}}
\DeclareMathOperator{\tr}{Tr}
\title[The CFHTLenS shear power spectrum]{A direct measurement of tomographic lensing power spectra from CFHTLenS}
\author[F. K\"ohlinger et al.]{F. K\"ohlinger$^{1, \, 2}$\thanks{E-mail:
fkoehlin@strw.leidenuniv.nl}, M. Viola$^1$, W. Valkenburg$^2$, B. Joachimi$^3$, H. Hoekstra$^1$ and K. Kuijken$^1$\\
$^{1}$Leiden Observatory, Leiden University, PO Box 9513, Leiden, NL-2300 RA, the Netherlands\\
$^{2}$Instituut--Lorentz, Leiden University, PO Box 9506, Leiden, NL-2300 RA, the Netherlands\\
$^{3}$Department of Physics and Astronomy, University College London, Gower Street, London WC1E 6BT, UK}
\begin{document}

\date{Accepted 0000 XXXX 00. Received 0000 XXXX 00; in original form 2015 XXXX 00}

\pagerange{\pageref{firstpage}--\pageref{lastpage}} \pubyear{2015}

\maketitle

\label{firstpage}

\begin{abstract}
We measure the weak gravitational lensing shear power spectra and their cross-power in two photometric redshift bins from the Canada--France--Hawaii Telescope Lensing Survey (CFHTLenS). The measurements are performed directly in multipole space in terms of adjustable band powers. For the extraction of the band powers from the data we have implemented and extended a quadratic estimator, a maximum likelihood method that allows us to readily take into account irregular survey geometries, masks, and varying sampling densities.
We find the 68 per cent credible intervals in the $\sigma_8$--$\Omega_{\rm m}$-plane to be marginally consistent with results from \textit{Planck} for a simple five parameter $\Lambda$CDM model. For the projected parameter $S_8 \equiv \sigma_8(\Omega_{\rm m}/0.3)^{0.5}$ we obtain a best-fitting value of $S_8 = 0.768_{-0.039}^{+0.045}$. This constraint is consistent with results from other CFHTLenS studies as well as the Dark Energy Survey. Our most conservative model, including modifications to the power spectrum due to baryon feedback and marginalization over photometric redshift errors, yields an upper limit on the total mass of three degenerate massive neutrinos of $\Sigma m_\nu < 4.53 \, {\rm eV}$ at 95 per cent credibility, while a Bayesian model comparison does not favour any model extension beyond a simple five parameter $\Lambda$CDM model. Combining the shear likelihood with \textit{Planck} breaks the $\sigma_8$--$\Omega_{\rm m}$-degeneracy and yields $\sigma_8=0.818 \pm 0.013$ and $\Omega_{\rm m} = 0.300 \pm 0.011$ which is fully consistent with results from \textit{Planck} alone.   
\end{abstract}

\begin{keywords}
cosmology: observations -- large-scale structure of Universe -- cosmological parameters -- gravitational lensing: weak.
\end{keywords}

\section{Introduction}
\label{sec:intro} 
The physical nature of the major components of current cosmological models is still unknown. Nevertheless, a simple six parameter model including dark matter and dark energy -- the $\Lambda$ dominated cold dark matter model ($\Lambda$CDM) -- has been proven very successful in explaining a multitude of cosmological observations ranging from the radiation of the cosmic microwave background (CMB, e.g. \citealt{Planck2015_CP}) to supernovae (e.g. \citealt{Riess2011}) and large-scale structure (LSS) probes (e.g. \citealt{Aubourg2014}).

The energy densities of dark matter and dark energy, at present, are very well constrained by the aforementioned observations. The next frontier is pinning down the evolution of both dark species, and to observe effects from massive neutrinos. One promising probe is the growth of structure as inferred from cosmic shear: the (very) weak lensing effect due to cosmic large-scale structure bending the light perpendicular to the line-of-sight between observer and background galaxies according to Einstein's equivalence principle. The coherent image distortions -- the shear -- due to the gravitational potential of a deflector can only be measured statistically, which requires averaging over large numbers of sources. Therefore, wide-field surveys covering increasingly larger volumes on the sky are required in order to improve the precision of the measurements. An analysis of the weak-lensing signal as a function of redshift is sensitive to the growth of structure, and is thereby indirectly sensitive to the expansion rate of the Universe as well as to the clustering behaviour of various matter species: massive neutrinos, dark energy, cold dark matter, etc.

In order to constrain the dark energy equation-of-state and its possible time evolution it is hence crucial to measure the cosmic shear signal in different redshift slices \citep{Heymans2013, Benjamin2013, DES_cosmo2015} or directly in 3D \citep{Kitching2014}.

Massive neutrinos also leave their distinct physical imprints on the matter power spectrum and hence can be probed using weak lensing (e.g. \citealt{Lesgourgues2006} and references therein). Theoretically it is straightforward to study these features directly in Fourier space, i.e., in terms of shear-shear power spectra. Traditionally, lensing analyses employ real-space correlation functions for measuring cosmic shear. This introduces further complications in the comparison of observations with theory (cf. Section~4.3.2 of \citealt{Planck2015_DE}), because different scales are highly correlated. Hence, the signal at very non-linear scales requires proper modelling in order to avoid any bias in the cosmological parameters. This is generally challenging due to our limited understanding of the effect of baryons on the non-linear matter power spectrum (e.g. \citealt{Semboloni2011, Semboloni2013}). Therefore, in this paper we apply a method for extracting the data in multipole space and in different redshift bins in terms of band powers of the lensing power spectrum. In order to achieve this we have implemented and expanded the quadratic estimator method originally formulated in the context of weak lensing by \citet{Hu2001}. The first applications of this technique to measured shear data were presented in \citet{Brown2003} and \citet{Heymans2005} using the COMBO-17 and GEMS data sets, respectively. More recently, \citet{Lin2012} applied the quadratic estimator technique to shear data measured from the Sloan Digital Sky Survey (SDSS) Stripe 82. Other recent direct shear power spectrum analyses include the Dark Energy Survey (DES) \citep{Becker2015} analysis and the SDSS-FIRST cross-power spectrum analysis of \citet{Demetroullas2015}. All these studies did not split the power spectrum analysis into redshift bins yet and the latter two studies employed a pseudo-$C(\ell)$ power spectrum approach, the other major technique for direct power spectrum measurements. \citet{Alsing2015} recently presented a hierarchical inference method that also makes direct use of the shear power spectrum. 

In this paper we apply our expanded tomographic version of the quadratic estimator to publicly available data from the Canada--France--Hawaii Telescope Lensing Survey (CFHTLenS, \citealt{Heymans2012}). CFHTLenS is currently the statistically most constraining weak lensing data set and covers an area of about $154$ square degrees on the sky. The data include also photometric redshifts which thus allows us to carry out a tomographic analysis. As a further benefit to the state-of-the-art data, CFHTLenS has already been used before in cosmological analyses \citep{Heymans2013, Benjamin2013, Kilbinger2013, Kitching2014} which enables us to directly cross-check our results with the literature. 

The paper is structured as follows: in Section~\ref{sec:theory} we introduce the weak lensing formalism in terms of power spectra. In Section~\ref{sec:qe} we describe the theory of the quadratic estimator approach and generalize it to include tomography. Section~\ref{sec:data_meas} provides a brief overview of the CFHTLenS data and how to perform shear measurements with it. Before presenting the extracted lensing power spectra in Section~\ref{sec:signal}, we test and validate the method on mock data in Section~\ref{sec:mocks}. From the shear power spectra we derive cosmological parameters and discuss our results in Section~\ref{sec:cosmo_inference}. Finally we present our conclusions in Section~\ref{sec:conclusions}.

\section{Theory}
\label{sec:theory}
 
The deflection of light due to mass is a consequence of Einstein's principle of equivalence and is termed gravitational lensing. One particular case of gravitational lensing is weak lensing, the very weak but coherent image distortions of background sources due to the gradients of the gravitational potential of a deflector in the foreground. 

These image distortions can only be measured in a statistical sense, given the fact that galaxies are intrinsically elliptical, by averaging over large numbers of background sources. The resulting correlations in the galaxy shapes can be used to study the evolution of all the intervening large-scale structure between the sources and the observer, in that sense the whole Universe acts as a lens. This particular form of weak lensing is called cosmic shear and studied best in terms of wide-field surveys covering increasingly more volume in the sky (cf. \citealt{Kilbinger2015} for a recent review). We intentionally skip a more basic, mathematical introduction of gravitational lensing and weak lensing in particular and refer the reader for details on that to the standard literature (e.g. \citealt{BartelmannSchneider2001}).

A wide-field observation of the sky as part of a weak lensing survey yields two main observables: the ellipticity of galaxies and their (photometric) redshifts. The estimates of the ellipticity components $e_{1}$, $e_{2}$ at angular positions $\bmath{n}_i$ can be binned into pixels $i = 1,\, ..., \, N_{pix}$ and (photometric) redshift bins $z_\mu$. The averages of the measured ellipticities in each pixel are unbiased estimates of the two components of the spin-2 shear field, $\gamma_1(\bmath{n}, \, z_\mu)$ and $\gamma_2(\bmath{n}, \, z_\mu)$, which is sourced by the convergence field $\kappa$. In the limit of the flat-sky approximation the Fourier decomposition of this field can be expressed as
\begin{equation}
\begin{split}
\gamma_1(\bmath{n}, \, z_\mu)\pm \rm{i} \gamma_2(\bmath{n}, \, z_\mu) = \int & \frac{\diff^2 \ell}{(2{\rm \pi})^2} W(\bmath{\ell}) \\
& \times [\kappa(\bmath{\ell}, \, z_\mu) \pm \rm{i} \beta(\bmath{\ell}, \, z_\mu)] \\
& \times e^{\pm 2\rm{i} \varphi_\ell} e^{\rm{i} \bmath{\ell}\cdot\bmath{n}} \, ,
\end{split}
\end{equation}
where $\varphi_\ell$ is the angle between the two-dimensional vector $\bmath{\ell}$ and the $x$-axis. To first order for the lensing of density perturbations the field $\beta$ vanishes in the absence of any systematics. However, we still want to measure it as a systematic test and therefore include it in our notation. 
The Fourier transform of the pixel window function is denoted as $W(\bmath{\ell})$. This function can explicitly be written out for square pixels of side length $\sigma_{\mathrm{pix}}$ in radians as
\begin{equation}
W(\bmath{\ell}) = j_0 \left(\frac{\ell_x \sigma_{\mathrm{pix}}}{2} \cos \varphi_\ell \right) j_0 \left(\frac{\ell_y \sigma_{\mathrm{pix}}}{2} \sin \varphi_\ell \right) \, ,
\end{equation}
where the zeroth-order spherical Bessel function is defined as $j_0(x)=\sin(x)/x$.\\

The two-point statistics of the shear field can either be expressed in real-space correlation functions or equivalently in terms of their Fourier transforms, the shear power spectra.

Following \citet{Hu2001} and expanding the notation to also include tomographic bins we write out the shear correlations between pixels $\bmath{n}_i$ and $\bmath{n}_j$ in terms of their power spectra as:
\begin{align} \label{eq:shear_corr}
\langle \gamma_{1i\mu} \gamma_{1j\nu} \rangle & = \int \frac{\diff^2 \ell}{(2{\rm \pi})^2} [C^{\mathrm{EE}}_{\mu\nu}(\ell)\cos^2 2\varphi_\ell \nonumber \\
& \hphantom{{} = \int} + C^{\mathrm{BB}}_{\mu\nu}(\ell) \sin^2 2\varphi_\ell  \nonumber \\
& \hphantom{{} = \int} -C^{\mathrm{EB}}_{\mu\nu}(\ell) \sin 4  \varphi_\ell]W^2(\bmath{\ell})e^{\rm{i}\bmath{\ell}\cdot(\bmath{n}_i-\bmath{n}_j)} \, , \nonumber \\
\langle \gamma_{2i\mu} \gamma_{2j\nu} \rangle & = \int \frac{\diff^2 \ell}{(2{\rm \pi})^2} [C^{\mathrm{EE}}_{\mu\nu}(\ell)\cos^2 2\varphi_\ell \nonumber \\
& \hphantom{{} = \int} +C^{\mathrm{BB}}_{\mu\nu}(\ell) \sin^2 2\varphi_\ell \nonumber \\ 
& \hphantom{{} = \int} +C^{\mathrm{EB}}_{\mu\nu}(\ell) \sin 4 \varphi_\ell]W^2(\bmath{\ell})e^{\rm{i}\bmath{\ell}\cdot(\bmath{n}_i-\bmath{n}_j)} \, , \nonumber \\
\langle \gamma_{1i\mu} \gamma_{2j\nu} \rangle & = \int \frac{\diff^2 \ell}{(2{\rm \pi})^2} [\tfrac{1}{2} \, (C^{\mathrm{EE}}_{\mu\nu}(\ell) - C^{\mathrm{BB}}_{\mu\nu}(\ell)) \sin 4\varphi_\ell \nonumber \\ 
& \hphantom{{} = \int} +C^{\mathrm{EB}}_{\mu\nu}(\ell) \cos 4 \varphi_\ell ] W^2(\bmath{\ell})e^{\rm{i}\bmath{\ell}\cdot(\bmath{n}_i-\bmath{n}_j)} \, ,
\end{align}
where we have suppressed the arguments of the shear components $\gamma_a(\bmath{n}_i, \, z_\mu)$ for clarity.

In the absence of systematic errors and shape noise the cosmological signal is contained in the E-modes and their power spectrum is equivalent to the convergence power spectrum, i.e.,  $C^{\mathrm{EE}}(\ell)=C^{\kappa\kappa}(\ell)$ and $C^{\mathrm{BB}}(\ell)=0=C^{\mathrm{EB}}(\ell)$. Shot noise will generate equal power in E- and B-modes.

The E-mode or convergence power spectra can be predicted for a given cosmological model:
\begin{equation}
\label{eq:theo_power_spec}
C_{\mu \nu}^{\mathrm{EE}}(\ell) = \frac{9\Omega^{2}_{\rm m}H_0^4}{4c^4} \int_{0}^{\chi_{\rm H}} \diff \chi \, \frac{g_\mu(\chi) g_\nu(\chi)}{a^2(\chi)} P_\delta\left(k=\frac{\ell}{f_\mathrm{K}(\chi)}; \chi \right) \, ,
\end{equation}
where $\chi$ is the radial comoving distance, $\chi_{\rm H}$ the distance to the horizon, $a(\chi)$ the scale factor, $P_\delta(k; \chi)$ is the three-dimensional matter power spectrum, and the angular diameter distance is denoted as $f_\mathrm{K}(\chi)$. Note that we use the Limber approximation \citep{Limber1954} in the equation above and the indices $\mu$, $\nu$ run over the tomographic bins.

The lensing kernels $g_\mu(\chi)$ are a measure for the lensing efficiency in each tomographic bin $\mu$ and can be written as
\begin{equation}
\label{eq:lensing_kernel}
g_\mu(\chi) = \int_\chi^{\chi_\mathrm{H}} \diff \chi^{\prime} \, n_\mu(\chi^{\prime})\frac{f_\mathrm{K}(\chi^{\prime}-\chi)}{f_\mathrm{K}(\chi^{\prime})} \, ,
\end{equation}
where $n_\mu(\chi) \, \diff \chi=p_\mu(z) \, \diff z$ is the source redshift distribution.

\section{Quadratic Estimator} 
\label{sec:qe}

We summarize here the method originally proposed by \citet{Hu2001} but make also extensive use of the summary provided by \citet{Lin2012}. Furthermore, we generalize the approach to include tomographic redshift bins.  

We start by assuming that the likelihood of the measured shear field in terms of band powers $\bmath{\mathcal{B}}$ is Gaussian over most scales of interest for our analysis, i.e.,
\begin{equation}
\mathcal{L} = \frac{1}{(2{\rm \pi})^N|\bmath{{\rm C}}(\bmath{\mathcal{B}})|^{1/2}}\exp{ \left[-\tfrac{1}{2} \, \bmath{d}^T[\bmath{{\rm C}}(\bmath{\mathcal{B}})]^{-1} \bmath{d} \right]} \, ,
\end{equation}
where $\bmath{d}$ denotes the data vector with components
\begin{equation}
d_{\mu ai} = \gamma_{a}(\bmath{n}_i, \ z_\mu) \, .
\end{equation}
It contains the two components of the measured shear in each pixel $\bmath{n}_i$ per redshift bin $z_\mu$ (note that the indices are all interchangeable as long as the order is consistent throughout the algorithm below). The full covariance matrix $\bmath{{\rm C}}$ is the sum of the cosmological signal $\bmath{{\rm C}}^{{\rm sig}}$ and the noise $\bmath{{\rm C}}^{{\rm noise}}$. The latter includes the contribution from shape and measurement errors.
We use the set of equation~\ref{eq:shear_corr} to build up the lensing signal correlation matrix, where we label the shear components with indices $a, \ b$, pixels with indices $i, \ j$, and redshift bins with indices $\mu, \ \nu$:
\begin{equation}
\label{eq:sig_mat}
\bmath{{\rm C}}^{\mathrm{sig}}   = \langle \gamma_a(\bmath{n}_i, \, z_\mu) \gamma_b(\bmath{n}_j, \, z_\nu) \rangle \, .
\end{equation}
Furthermore, the contribution of shape noise to the signals can be encoded in the matrix 
\begin{equation} \label{eq:noise}
\bmath{{\rm C}}^{\mathrm{noise}} = \frac{\sigma_{\gamma}^2}{N_{i\mu}} \delta_{ij} \delta_{ab} \delta_{\mu\nu} \, , 
\end{equation}
where $\sigma_{\gamma}$ denotes the root-mean-square intrinsic ellipticity per ellipticity component for all the galaxies and $N_{i\mu}$ is the effective number of galaxies per pixel $i$ in redshift bin $z_\mu$.\footnote{The effective number of galaxies per pixel can be calculated using equation~\ref{eq:n_eff} multiplied by the area of the pixel $\Omega$.} Thus we assume that shape noise is neither correlated between different pixels $\bmath{n}_i$, $\bmath{n}_j$, and shear components $\gamma_a$, $\gamma_b$, nor between different redshift bins $z_\mu, \ z_\nu$. This is a well-motivated assumption as long as the pixel noise of the detector is uncorrelated. 

We approximate the angular power spectra $C^{\theta}_{\mu\nu}(\ell)$ with piece-wise constant band powers $\mathcal{B}_{\zeta\theta\beta}(\ell)$ of type $\theta \in ({\rm EE}, {\rm BB}, {\rm EB})$ spanning a range of multipoles $\ell$ within the band $\beta$. The index $\zeta$ runs only over \textit{unique} redshift bin correlations. This enables us to write the components of the cosmic signal covariance matrix as a linear combination of these band powers:
\begin{equation} \label{eq:signal}
\begin{split}
C^{{\rm sig}}_{(\mu\nu)(ab)(ij)} = &\sum_{\zeta, \theta, \beta} \mathcal{B}_{\zeta\theta\beta} M_{\zeta(\mu\nu)}\int_{\ell \in \beta} \frac{\mathrm{d}\ell}{2(\ell+1)} \\
& \times \left[ w_0(\ell)I^{\theta}_{(ab)(ij)} + \frac{1}{2} \, w_4(\ell) Q^{\theta}_{(ab)(ij)} \right] \, .
\end{split}
\end{equation} 
The term in brackets in the above equation encodes the geometry of the shear field including masks and its decomposition in Fourier space. The matrices $\bmath{\rm M}_\zeta$ map the redshift bin indices $\mu, \nu$ to the unique correlations $\zeta$ possible between those: for $n_z$ redshift bins there are only $n_z(n_z+1)/2$ unique correlations because $z_\mu \times z_\nu = z_\nu \times z_\mu$. The explicit expressions for these matrices and the matrices $\bmath{{\rm I}}^{\theta}$ and $\bmath{{\rm Q}}^{\theta}$ are given in Appendix~\ref{app:indices}.

The best-fitting band powers $\mathcal{B}_{\zeta\theta\beta}$ are determined by finding the cosmic signal $\bmath{{\rm C}}^{{\rm sig}}$ which describes the measured shear data the best. For that purpose we use the Newton--Raphson method iteratively in order to find the root of $\diff \mathcal{L}/ \diff \mathcal{B}_{A} = 0$ \citep{Bond1998, Seljak1998}. An improved estimate for the band powers $\mathcal{B}_A$ is found by evaluating the expression 
\begin{equation}
\label{eq:stepping}
\delta \mathcal{B}_{A} \propto \sum_B \tfrac{1}{2}(\bmath{{\rm F}}^{-1})_{AB} \tr [(\bmath{d}\bmath{d}^T - \bmath{{\rm C}})(\bmath{{\rm C}}^{-1}\bmath{{\rm D}}_{A}\bmath{{\rm C}}^{-1}) ] \, ,
\end{equation} 
where we have introduced now the super-index $A$ for a particular index combination $(\zeta\theta\beta)$. The matrices $\bmath{{\rm D}}_{A}$ are derivatives of the full covariance matrix with respect to any band power combination. We skip here a rigorous definition of $\bmath{{\rm D}}_{A}$ and refer the reader to Appendix~\ref{app:indices} for derivations of these expressions. The elements of the Fisher matrix $\bmath{{\rm F}}$ can be calculated as \citep{Hu2001}
\begin{equation}
\label{eq:Fisher}
F_{AB} = \tfrac{1}{2} \tr(\bmath{{\rm C}}^{-1}\bmath{{\rm D}}_{A}\bmath{{\rm C}}^{-1}\bmath{{\rm D}}_{B}) \, .
\end{equation} 
In previous work (cf. \citealt{Hu2001, Lin2012}), the inverse of the Fisher matrix was used as an estimator of the covariance between the extracted band powers. We refrain from following this approach since the inverse Fisher matrix is only an approximation of the true covariance in the Gaussian limit. Hence, we decided to estimate the covariance of the band powers from mock data instead. We present a detailed discussion of this approach in Sec.~\ref{sec:covariance}.

For the comparison of the measured band powers to theoretical predictions, we have to take into account that each measured band power $\mathcal{B}_{A} = \mathcal{B}_{\zeta\theta\beta}$ samples the power spectra with its own window function. This can be computed by noting that the expectation value of the band power, $\langle \mathcal{B}_{\zeta\theta\beta} \rangle$, is related to the power spectrum at each wave number $\mathcal{B}_{\zeta\theta}(\ell) = \ell (\ell+1) C_{\zeta\theta}(\ell)/(2{\rm \pi})$ through the band power window function $W_{\zeta\theta\beta}(\ell)$ \citep{Knox1999, Lin2012}, i.e.,
\begin{equation} \label{eq:conv_window_func}
\langle \mathcal{B}_{\zeta\theta\beta} \rangle = \sum_\ell W_{\zeta\theta\beta}(\ell) \mathcal{B}_{\zeta\theta}(\ell) \, , 
\end{equation}
where the sum is calculated for integer multipoles $\ell$\footnote{For the cosmological analysis we employ a range $80 \leq \ell \leq 2600$. The lower limit is set by the smallest multipole $\ell$ included in the analysis and the upper limit must include multipoles $\ell$ higher than the maximum $\ell$ used in the analysis (cf. Section~\ref{sec:data_meas}).}. The elements of the window function matrix can be derived as \citep{Lin2012}
\begin{equation} \label{eq:window}
W_{\zeta\theta\beta}(\ell) = \sum_{\chi, \eta, \lambda} \tfrac{1}{2}(\bmath{{\rm F}}^{-1})_{(\zeta\theta\beta)(\chi\eta\lambda)} T_{\chi\eta\lambda}(\ell) \, ,
\end{equation} 
where $\bmath{{\rm F}}^{-1}$ denotes the inverse of the Fisher matrix (cf. equation~\ref{eq:Fisher}). The trace matrix $\bmath{{\rm T}}$ is defined as
\begin{equation} \label{eq:trace}
T_{\zeta\theta\beta}(\ell) = \tr(\bmath{{\rm C}}^{-1}\bmath{{\rm D}}_{\zeta\theta\beta}\bmath{{\rm C}}^{-1}\bmath{{\rm D}}_{\ell}) \, .
\end{equation}
The derivative $\bmath{{\rm D}}_{\ell}$ denotes the derivative of the full covariance $\bmath{\rm C}$ with respect to the power at a single multipole $\ell$. We write it out explicitly in Appendix~\ref{app:indices} (cf. equation~\ref{eq:deriv_BWM}).

The likelihood-based quadratic estimator automatically accounts for any irregularity in the survey geometry or data sampling while it still maintains an optimal weighting of the data. This is important when dealing with real data because it allows for employing sparse sampling techniques and it can deal efficiently with (heavily) masked data. The whole method and in particular its ability to deal with masks are tested extensively in Section~\ref{sec:mocks} before we apply it to data from CFHTLenS in Section~\ref{sec:signal}.

\section{CFHTLenS measurements}
\label{sec:data_meas}

In the following analysis we use the publicly available data\footnote{\url{http://www.cfhtlens.org/astronomers/data-store}} from the lensing analysis of the Canada--France--Hawaii Legacy Survey, hereafter referred to as CFHTLenS \citep{Heymans2012}. 
The survey consists of four patches (W1, W2, W3, W4) covering a total area of $\approx 154 \deg^2$. Due to stellar haloes or artifacts in the images 19 per cent of the area is masked. The lensing data we use in this work are a combination of data processing with {\small THELI} \citep{Erben2013}, shear measurements with lens\emph{fit} \citep{Miller2013}, and photometric redshift measurements with PSF-matched photometry \citep{Hildebrandt2012}. A full systematic error analysis of the shear measurements in combination with the photometric redshifts is presented in \citet{Heymans2012}, with additional error analyses of the photometric redshift measurements presented in \citet{Benjamin2013}. One of the main results of those extensive systematic tests was the rejection of 25 per cent of the CFHTLenS tiles ($1 \deg^2$ each) for cosmic shear studies. In this work we only use the 75 per cent of the tiles which passed the systematic tests as outlined in \citet{Heymans2012}. Note that this causes considerable large scale masking in each patch. 

Photometric redshift measurements have also been extensively tested \citep{Hildebrandt2012, Benjamin2013} and they were found reliable in the range $0.1<Z_{\rm B}<1.3$, where $Z_{\rm B}$ is the peak of the photometric redshift posterior distribution as computed by {\small BPZ} \citep{Benitez2000}. In our analysis we only use galaxies in this redshift range.

We compile all tiles associated to a particular CFHTLenS patch into a single shear catalogue. Coordinates in these catalogues are given in right ascension $\alpha$ and declination $\delta$ of a spherical coordinate system. We deproject these spherical coordinates into flat coordinates via a tangential plane projection. We centre the projection, its tangent point, on the central pointing of each patch. In order to measure shears from the ellipticity components $e_1$, $e_2$ as measured by lens\emph{fit}, we first divide the deprojected patch into square pixels of side length $\sigma_{\rm pix}$. We estimate the shear components $g_a$ per pixel at position $\bmath{n}=(x_c, y_c)$ from the ellipticity components $e_a$ inside that pixel:
\begin{equation}
g_a(x_c, y_c) = \frac{\sum_{i} w_i (e_{a, i}-c_{a, i})}{(1+m)\sum_{i} w_i} \, ,
\end{equation}
where the index $i$ runs over all objects inside the pixel and the index $a$ is either $1$ or $2$ for the two shear and ellipticity components, respectively. The weights $w$ are computed during the shape measurement with lens\emph{fit} and they account both for the intrinsic shape noise and measurement errors. The subscript of the coordinates indicates that the position of the average shear is taken to be at the centre of the pixel. Note that we assume the galaxies are distributed uniformly in the shear pixels. Although this is a simplifying assumption we argue that it has only minor effects in the measurement considering the general width of the band powers. We define distances $r_{ij}=|\bmath{n}_i-\bmath{n}_j|$ and angles $\varphi=\arctan{(\Delta y/\Delta x)}$ between all pixels $i$, $j$ which enter eventually in the quadratic estimator algorithm (cf. Section~\ref{sec:qe} and Appendix~\ref{app:indices}). 

In each pixel we apply an average multiplicative correction $(1+m)$ to the measured shear. This is necessary because of noise bias in shear measurements \citep{Melchior2012, Refregier2012, Miller2013}. The multiplicative correction has been computed from a dedicated suite of image simulation mimicking CFHTLenS data \citep{Miller2013}. Moreover, we apply to each measured ellipticity an additive correction $c_a$ which is computed from all the pass-tiles by requiring that the average ellipticity must vanish across the survey as a function of galaxy size and signal-to-noise \citep{Heymans2012}. For CFHTLenS $c_1$ was found to be zero but for $c_2$ a correction per object has to be applied \citep{Heymans2012}.

The highest multipole $\ell_{\rm pix}$ up to which we want to extract band powers employing the quadratic estimator method (cf. Section~\ref{sec:qe}) is on the one hand set by the scales we want to investigate because of expected modifications due to baryon feedback or massive neutrinos (cf. Section~\ref{sec:theo_ps}). On the other hand the simplifying assumptions of the algorithm such as Gaussianity also limit the maximum $\ell_{\rm pix}$. Hence, we only probe into the mildly non-linear regime and consider a multipole $\ell_{\rm pix} \approx 2400$ as the maximal physical scale resolved. This corresponds to an angular scale of $0.15 \degr = 9 \arcmin$ and thus sets the pixel size $\sigma_{\rm pix}$. We keep parameters fixed throughout all CFHTLenS patches such as the side length of the shear pixels, $\sigma_{\rm pix}$, measured intrinsic shape noise per ellipticity component, $\sigma_\gamma = 0.279$, and band power intervals. Because the sizes of the CFHTLenS patches are very different, the largest distance between shear pixels differs. Therefore, we limit our analysis to $\ell_{\rm field} \geq 80$ (corresponding to an angular separation of pixels of about $ \sim 4.5 \degr$), but note that even lower multipoles suffer from more sample variance. In summary, the physical scales for our analysis are $80 \leq \ell \leq 2300$, which corresponds to angular scales $0.15 \degr \leq \vartheta \leq 4.5 \degr$. In total, we choose seven band power intervals enclosing these physical scales as shown in Table~\ref{tab:bp_intervals} for the E-mode signal extraction. The width of each band should at least be two times as wide as $\ell_{\rm field}$ in order to minimize correlations between the bands \citep{Hu2001}. The band powers for the B-mode signal extraction are the same except that we omit the lowest band power. Note that the first band power includes scales below $\ell_{\rm field}$ intentionally in order to absorb any DC offsets in the data. The last band should include multipoles above $\ell_{\rm pix}$, because the window function of square pixels has a long tail to high multipoles. In that sense the enclosing bands are designed to catch noise and therefore they are dropped in the cosmological analysis.
\begin{table}
	\caption{Band power intervals}
	\label{tab:bp_intervals}
	\begin{center}
		\begin{tabular}{ c c c c }
			\hline
			Band No.& $\ell$--range& $\vartheta$--range& Comments\\
			\hline
			1& 30--80& $720\arcmin$--$270\arcmin$& a), b)\\
			2& 80--260& $270\arcmin$--$83\arcmin$& --\\
			3& 260--450& $83\arcmin$--$48\arcmin$& --\\
			4& 450--670& $48\arcmin$--$32\arcmin$& --\\
			5& 670--1310& $32\arcmin$--$16\farcm5$& --\\
			6& 1310--2300& $16\farcm5$--$9\farcm4$& a)\\
			7& 2300--5100& $9\farcm4$--$4\farcm2$& a)\\
			\hline
		\end{tabular}
	\end{center}
	\medskip 
	\textit{Notes.} a) Not used in cosmological analysis. b) No B-mode extracted. \\
	The $\vartheta$-ranges are just an indication and cannot be compared directly to $\vartheta$-ranges used in real-space correlation function analyses due to the non-trivial functional dependence of these analyses on Bessel functions.
\end{table}
We compute the effective number density of galaxies that is used in the lensing analysis and in the creation of mock data (cf. Section~\ref{sec:mocks}) following the definition of \citet{Heymans2012}: 
\begin{equation}
\label{eq:n_eff}
n_\mathrm{eff} = \frac{1}{\Omega}\frac{(\sum_i w_i)^2}{\sum_i w_i^2} \, ,
\end{equation} 
where $\Omega$ is the unmasked area used in the analysis and $w$ is again the lens\emph{fit} weight. We show all effective number densities per patch and redshift bin in Table~\ref{tab:n_eff}. 
\begin{table}
	\caption{Effective number densities}
	\label{tab:n_eff}
	\begin{center}
		\begin{tabular}{ c c c c c }
			\hline
			redshift bin& W1& W2& W3& W4\\
			\hline
			$z_1$: $0.50 < Z_{\rm B} \leq 0.85$& 3.36& 2.80& 3.48& 3.25\\
			$z_2$: $0.85 < Z_{\rm B} \leq 1.30$& 2.86& 2.00& 2.63& 2.22\\
			\hline
		\end{tabular}
	\end{center}
	\medskip 
	\textit{Notes.} Shown is the effective number density of galaxies $n_\mathrm{eff}$ (cf. equation~\ref{eq:n_eff}) in arcmin$^{-2}$ for all four CFHTLenS patches per tomographic redshift bin used in this analysis.
\end{table}

Following the conclusions from \citet{Benjamin2013} regarding intrinsic galaxy alignments, which we discuss in more detail in Section~\ref{sec:signal}, we define two redshift bins $z_1$ and $z_2$ in the ranges $z_1: 0.50 < Z_{\rm B} \leq 0.85$ and $z_2: 0.85 < Z_{\rm B} \leq 1.30$. These cuts are performed with respect to the peak of each galaxy's photometric redshift distribution $Z_{\rm B}$. For each of the two tomographic bins we compute the galaxy redshift distribution by summing the posterior photometric redshift distribution of all galaxies in the bin, weighted by the lens\emph{fit} weight:
%However, the galaxies in the lensing catalogue also contain the full photometric redshift probability density distribution $p(z)$. From these individual photometric redshift distributions we calculate the full photometric redshift distribution of a whole patch per tomographic bin:
\begin{equation}
p(z) = \frac{\sum_i w_i p_i(z)}{\sum_i w_i} \, . 
\end{equation}
The full galaxy redshift distribution is required in the calculation of the theoretical lensing power spectrum (cf. equation~\ref{eq:lensing_kernel}) and it is also needed in the creation of additional mock data (cf. Section~\ref{sec:mocks}).

\section{Method validation and covariances}
\label{sec:mocks}

In order to test and validate the algorithm outlined in Section~\ref{sec:qe} we employ two types of mock data: first we make use of the publicly available CFHTLenS Clone\footnote{\url{http://vn90.phas.ubc.ca/jharno/CFHT_Mock_Public/}} \citep{Heymans2012} and second we use Gaussian random fields (GRFs). This twofold approach is necessary since the multipole scales we employ in the cosmological analysis of Section~\ref{sec:cosmo_inference} are not covered in the CFHTLenS Clone.  

The CFHTLenS Clone is a mock galaxy catalogue that consists of 184 independent line-of-sight shear (and convergence) maps with a side length of $\approx 3.58 \degr$. These were extracted via ray-tracing through the TCS simulation suite \citep{Harnois2012} which was produced with the {\small CUBEP$^3$M} N-body code \citep{Harnois2013}. The CFHTLenS Clone is especially tailored to CFHTLenS in terms of the redshift distribution of lensing sources and the noise properties including, for example, realistic small scale masks (due to stars etc.). In addition to these small scale masks, we randomly mask out three non-overlapping tiles of $\approx 1 \deg^2$ each per shear field in order to mimic the effect of the additional `bad field' masks also employed in the data. These mask typically 25 per cent of the total area of a patch (cf. Section~\ref{sec:data_meas} and \citealt{Heymans2012}) and their distribution over a patch does not show any systematic preferences. The input cosmology used in the creation of the CFHTLenS Clone is WMAP5-like \citep{Komatsu2009} and summarized in Table~\ref{tab:input_cosmology}.
Eventually, we want to extract scales on the order of several degrees from the data. \citet{Kilbinger2013} showed, however, that the power on large scales is significantly underestimated in the CFHTLenS Clone.
\begin{table}
	\caption{Fiducial cosmology of the CFHTLenS Clone and the Gaussian random fields (GRFs)}
	\label{tab:input_cosmology}
	\begin{center}
		\begin{tabular}{ c c c c c c c}
			\hline
			$\Omega_\mathrm{m}$& $\Omega_\Lambda$& $\Omega_\mathrm{b}$& $h$& $n_\mathrm{s}$& $\sigma_8$& $\Sigma m_\nu$\\			
			\hline
			0.279& 0.721& 0.046& 0.701& 0.96& 0.817& $0 \, {\rm eV}$\\
			\hline
		\end{tabular}
	\end{center}
	\medskip 
	\textit{Notes.} Cosmological parameters used in the creation of the CFHTLenS Clone \citep{Heymans2012} which were also used to create the Gaussian random field (GRF) realizations.
\end{table}

In order to also validate the signal extraction on large scales, we created 184 Gaussian random field realizations (GRFs) of shear fields in two tomographic bins. The fields are $20 \times 20 \, {\rm deg}^2$ each and generated from convergence power spectra that have been computed for the same cosmology as the clone, using the measured redshift distributions of our two tomographic bins and the modified \texttt{halofit} version of \citet{Takahashi2012} for the non-linear contributions to the matter power spectrum. Source galaxies are placed randomly in the fields with an arbitrary but high enough density of $10 \, {\rm arcmin}^{-2}$ per tomographic bin, and the shears are linearly interpolated to these positions. We apply the mosaic masks of each CFHTLenS patch to all GRF realizations in turn, and also apply the patch-specific 'bad field' mask pattern masking about 25 per cent of the total area of a CFHTLenS patch. When we compile the actual input mock catalogues from the GRF shear fields, we also add shape noise by resampling the GRF shear from a Rayleigh-distribution with width $\sigma_{\gamma}=0.279$ as measured from the data. Furthermore, we randomly sample lens\emph{fit} weights from the corresponding tomographic data catalogues such that the effective number densities (cf. equation~\ref{eq:n_eff}) in the GRF mock catalogues match the ones in the data (cf. Table~\ref{tab:n_eff}). 

The (inverse) Fisher matrices calculated in the quadratic estimator algorithm (cf. Section~\ref{sec:qe}) are only an approximation of the true (inverse) covariance of the extracted band powers in the Gaussian limit. In the context of a cosmological interpretation of the band powers, however, additional non--Gaussian contributions due to the non-linear evolution of the underlying matter power spectrum are expected (cf. \citealt{Takada2009}). Hence, we will use our mock data also for estimating a more realistic band power covariance matrix.

\subsection{Signal extraction validation}
\label{sec:signal_validation}

\begin{figure*}
	\centering
	\includegraphics[width=180mm]{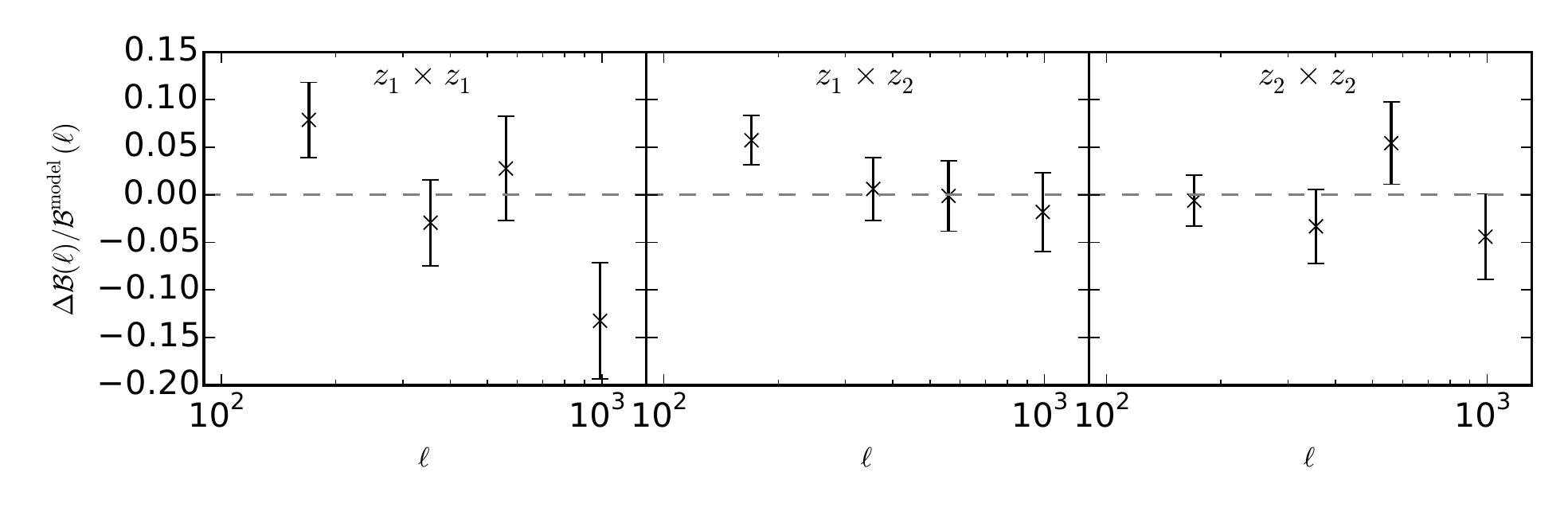}
	\caption{Residuals between the mean of measured E-mode band powers and predicted band powers for 184 Gaussian random field realizations of patch W3. The $1\sigma$-error on the mean includes the scaling by $1/\sqrt{N}$ for $N=184$ measurements. The predicted band powers use the known input cosmology (cf. Table~\ref{tab:input_cosmology}) and take the convolution with the band window function into account. The residuals of each redshift correlation are shown from left to right.}
	\label{fig:signals_grfs}
\end{figure*}

The input cosmology is known for the GRFs and the Clone, and we apply a realistic CFHTLenS mask to both sets of mock data. We extract the lensing power spectrum using the quadratic estimator from the GRFs and the Clone and compare it to the input power spectrum.
In Fig.\ref{fig:signals_grfs} we show the residuals between the mean of the extracted band powers and the predicted band powers for the input cosmology for patch W3. The $1\sigma$-errors on the mean include the scaling by $1/\sqrt{N}$ for $N=184$ GRFs for each tomographic bin correlation. The binning in multipoles $\ell$ is the same as the one we employ in the final data extraction (cf. Section~\ref{sec:data_meas} and Table~\ref{tab:bp_intervals}). Note that for this test we only extracted E-modes. For the calculation of the band power predictions we take the convolution with the band window matrices (cf. equation~\ref{eq:window}) into account but these are computed for only one randomly drawn realization of a GRF. This is due to long run-time and we have confirmed for patch W2 that the randomly drawn band window matrix is a fair representation of the ensemble (since the noise properties of all GRFs are very similar). Fig.~\ref{fig:signals_grfs} demonstrates that the quadratic estimator algorithm reproduces the input signal to sufficient accuracy and precision, especially given the actual noise level of the data (cf. Fig.~\ref{fig:signals_EE}). 
%Note, however, that we found at a later stage of the analysis an improved regularization scheme for the extraction of band powers. Employing this scheme is expected to lower the scatter in Fig.~\ref{fig:signals_grfs}.

\subsection{Band power covariance}
\label{sec:covariance}

The extracted band powers for each of the 184 shear fields from the Clone or 184 GRFs per patch can be used to estimate the run-to-run covariance of the band powers: 
\begin{equation}
\label{eq:covariance}
\bmath{\hat{{\rm C}}}_{\mathcal{B}(\ell)}(A, B) = \frac{1}{A_\mathrm{scale} (n_\mu-1)} \sum_{\mu}^{n_\mu}(\mathcal{B}_{A}^\mu-\mathcal{\bar{B}}_{A})(\mathcal{B}_{B}^\mu-\mathcal{\bar{B}}_{B}) \, ,
\end{equation} 
where $n_\mu$ is the total number of independent realizations per patch, $\mathcal{\bar{B}}$ is the mean of each band power per band over all realizations, $\mathcal{B}^{\mu}$ are the extracted band powers per realization, and $A_\mathrm{scale}$ is the scaling factor between each line-of-sight clone realization and the actual size of a CFHTLenS patch\footnote{Note that $A_\mathrm{scale}=1$ in the case of GRFs by construction. For the clones we follow \citep{Kilbinger2013} by matching 90 per cent (due to overlapping area between the tiles) of 16 CFHTLenS tiles (minus three due to the 'bad field' masking also employed in the clones) into one clone field. The ratio of this number over the number of used tiles in one patch (i.e. excluding the 'bad fields') is then $1/A_{\mathrm{scale}}$.}. The indices $A$ and $B$ denote again the previously introduced superindices and run over all bands and redshift correlations.

In order to combine the small-scale covariance estimated from the Clone and the large scale covariance based on the GRFs, we stitch both matrices together per patch by using the GRF covariance and then replacing all values associated with a band index for which we want to use the Clone covariance. Based on the extensive analysis of the Clone and the estimation of covariances from it in \citet{Kilbinger2013}, we decide to use values from the Clone covariance for multipoles $\ell \geq 670$ which corresponds to bands 5, 6, and 7 (cf. Table~\ref{tab:bp_intervals}). Note that bands 7, 6, and 1 are not included in any cosmological data analysis though (cf. Section~\ref{sec:cosmo_inference} and Table~\ref{tab:bp_intervals}).   

Due to noise the measured inverse covariance $\bmath{\hat{{\rm C}}}_{\mathcal{B}(\ell)}^{-1}$ is not an unbiased estimate of the true inverse covariance matrix $\bmath{{\rm C}}_{\mathcal{B}(\ell)}^{-1}$ \citep{Hartlap2007}. In order to derive an unbiased estimate of the inverse covariance we need to apply a correction derived in \citet{Kaufmann1967} so that $\bmath{{\rm C}}_{\mathcal{B}(\ell)}^{-1}=\alpha_{\rm K}\bmath{\hat{{\rm C}}}_{\mathcal{B}(\ell)}^{-1}$. Assuming a Gaussian distribution of the measured band powers $\mathcal{B}(\ell)$, this correction factor is:
\begin{equation}
\alpha_{\rm K} = \frac{n_\mu-p-2}{n_\mu-1} \, ,
\end{equation}
where $n_\mu$ is the total number of independent mocks, i.e. 184 in our case, and $p$ is the number of data points used in the analysis. In Section~\ref{sec:cosmo_inference} we combine the data of all four CFHTLenS patches consisting of four band powers in three tomographic power spectra per patch, thus $p=12$ for each `patch covariance'.

\begin{figure}
	\centering
	\includegraphics[width=84mm]{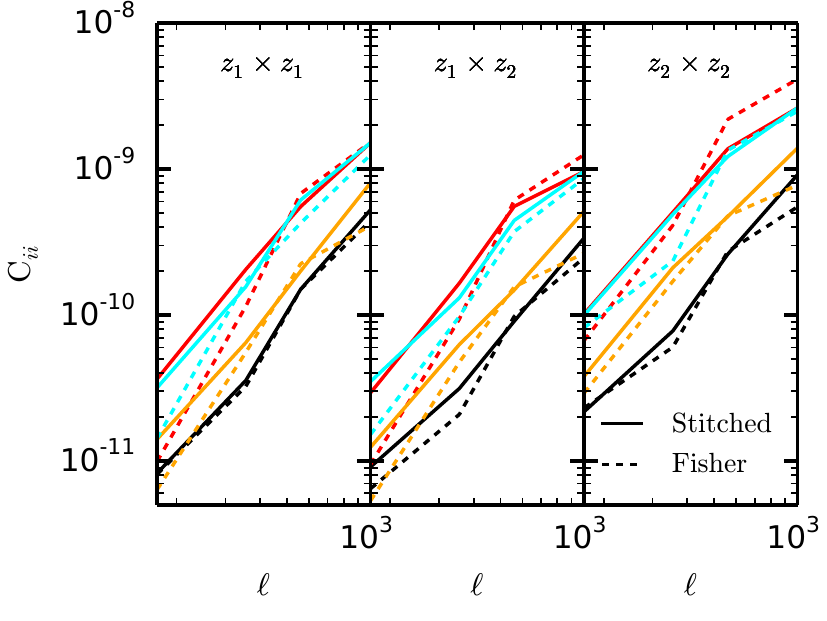}
	\caption{Comparison of correlation matrices for CFHTLenS patch W3: the stitched correlation matrix (upper right) is compared to the correlation matrix based on the inverse of the Fisher matrix (lower left; cf. equation~\ref{eq:Fisher}). We show only tomographic E-mode bins that enter in the final cosmological likelihood analysis, i.e., bins 0 to 3 correspond to $80 \leq \ell \leq 1310$ in the low redshift auto-correlation bin, bins 4 to 7 correspond to the same $\ell$--range in the redshift cross-correlation bin, and bins 8 to 11 correspond to the high redshift auto-correlation bin (cf. Tables ~\ref{tab:bp_intervals} and ~\ref{tab:n_eff}).}
	\label{fig:corr_matrices}
\end{figure}

\begin{figure}
	\centering
	\includegraphics[width=84mm]{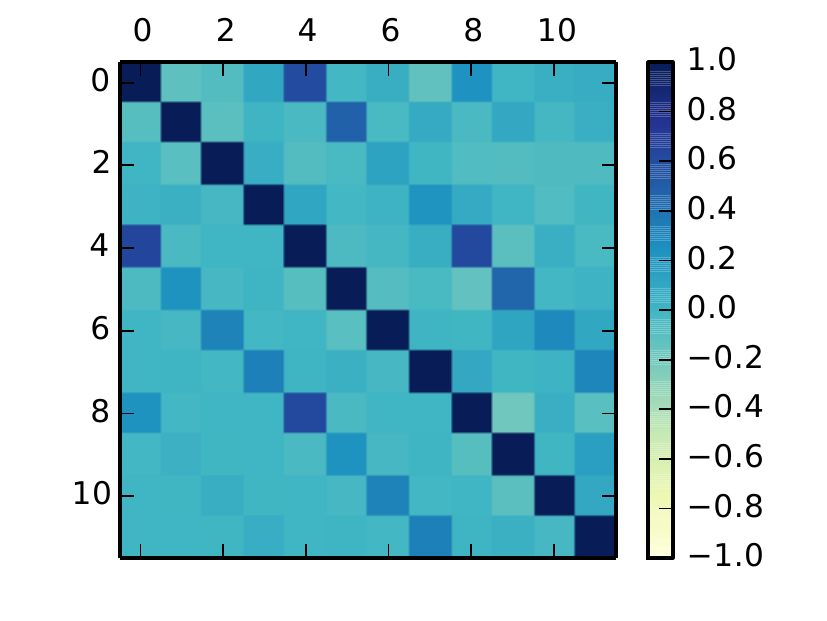}
	\caption{The variances calculated from the stitched covariance matrix (solid lines) and the inverse of the Fisher matrix (dashed lines) for all four CFHTLenS patches from bottom to top: W1 (black), W2 (red), W3 (orange), and W4 (cyan). From left to right we show the variances in the auto-correlation of the low redshift bin, in the cross-correlation between the low and the high redshift bin, and in the auto-correlation of the high redshift bin. We limit the $\ell$-range to the one considered in the cosmological analysis.}
	\label{fig:var_stitched_fisher}
\end{figure}

We compare the correlation matrix derived from the stitched covariance matrix with the correlation matrix based on the inverse Fisher matrix which is calculated in the quadratic estimator algorithm (cf. equation~\ref{eq:Fisher}) in Fig.~\ref{fig:corr_matrices} for patch W3. The correlation matrices are calculated by normalizing the corresponding covariance matrix with the factor $(\bmath{{\rm M}}_{AA}\, \bmath{{\rm M}}_{BB})^{-1/2}$, with $\bmath{{\rm M}}_{AB}=\bmath{{\rm C}}_{\mathcal{B}(\ell)}(A, B)$ or $\bmath{{\rm M}}_{AB}=\bmath{{\rm F}}_{AB}^{-1}$. We only include E-mode bands employed later in the cosmological analysis in this comparison and find that the matrix structure in both approaches is very similar albeit with the correlation based on the Fisher estimate being smoother, as expected. Finally, we compare both approaches in terms of their variance as shown in Fig.~\ref{fig:var_stitched_fisher} for each patch individually again only for E-modes used in the cosmological analysis. From this comparison we conclude that given the noise level in our data the Fisher approach still yields compatible error estimates. Nevertheless, we decide to use the stitched covariance for our subsequent analysis. This is also motivated by the fact that future surveys will yield significantly improved statistical noise levels and thus require a proper covariance estimation beyond the Fisher approach.  

\subsection{Computing resources}
\label{sec:computability}
We want to comment on the computational requirements for our tomographic quadratic estimator approach: the generalization of the method to include tomographic redshift bins is computationally demanding. The dimension of the covariance matrix defined in equation~\ref{eq:sig_mat} is set by the size of the shear field (times two for the two shear components) and the pixel scale. Introducing also two redshift bins increases the number of entries in this matrix by a factor of four. While this is still efficiently calculated in parallel for smaller patches like W2 ($\approx 22.6 \deg^2$) and W4 ($\approx 23.3 \deg^2$), it becomes demanding for patches W3 and W1 (e.g. ${\rm dim}(\bmath{\rm C}_{{\rm W}2}) = 3076^2$ versus ${\rm dim}(\bmath{\rm C}_{{\rm W}1}) = 9340^2$) even when exploiting multiprocessing and optimized libraries such as the Intel$^{\copyright}$ Math Kernel Library ({\small MKL}\footnote{Version number 11.0.4}). Nevertheless, the data extraction including the calculation of the band window matrices takes at most a day on typical cluster machines\footnote{24 cores @2.4 GHz, 256 GB RAM}. The computationally most demanding part in our current analysis, however, is the estimation of the covariance between the band powers. This required 184 runs on clones and 184 runs per GRF realization per patch. The total runtime for these calculations was on the order of a month on the same cluster configuration for one set of 184 realizations.

Ongoing and upcoming weak lensing surveys come with the advantage of at least an order of magnitude increase in survey area compared to CFHTLenS and more regular survey geometries. Therefore, it will be possible to split these surveys into a statistically meaningful number of patches still containing scales up to several degrees. This will allow for estimating the patch-to-patch covariance directly from the data via resampling techniques as an alternative to estimating it from mock data alone. However, this approach limits the lowest multipole scale to the patch-size and the run-to-run covariance will be underestimated at scales close to the patch-size. Finally, the rapid advance in terms of number of cores, clock speed, and internal memory of graphics processing units (GPUs) presents a solution to the increase in complexity when extending our approach to more redshift bins, and/or more band powers, and/or larger contiguous patch sizes. The advantage of GPUs lies in their customized design to solve linear algebra problems very efficiently and massively in parallel which meets exactly the requirements of the tomographic quadratic estimator approach. We leave an update and porting to GPU programming languages for future work.

\section{The CFHTLenS shear power spectrum}
\label{sec:signal}

For each of the four CFHTLenS patches we extract seven E-mode and six B-mode band powers enclosing an interval of physically interesting scales of $80 \leq \ell \leq 2300$ (cf. Section~\ref{sec:data_meas} and Table~\ref{tab:bp_intervals}). Moreover, we consider two broad mid- to high-redshift bins (cf. Table~\ref{tab:n_eff}) per CFHTLenS patch in order to perform a tomographic analysis following \citet{Benjamin2013}. Doing so, we attempt to decrease the expected contamination due to intrinsic galaxy alignments which is dominant at low redshifts and high multipoles $\ell$. \citet{Benjamin2013} concluded that any contamination due to intrinsic alignments is at most a few per cent for each redshift bin combination. We cross-check this conclusion with state-of-the-art intrinsic alignment models constrained by recent data from \citet{Sifon2015}. For the three intrinsic alignment models\footnote{These models include intrinsic alignment due to intrinsic ellipticity correlations, i.e. II, and also intrinsic alignment due to a gravitational shear-intrinsic ellipticity correlation, i.e. GI.} employed in there we do not find a significant contribution of intrinsic alignments to the cosmological signal in any of the redshift bin correlations and $\ell$-scales employed in our subsequent cosmological analysis. Based on these results intrinsic alignments will be ignored in the modelling of the signal in our subsequent analysis.
\begin{figure*}
	\centering
	\includegraphics[width=180mm]{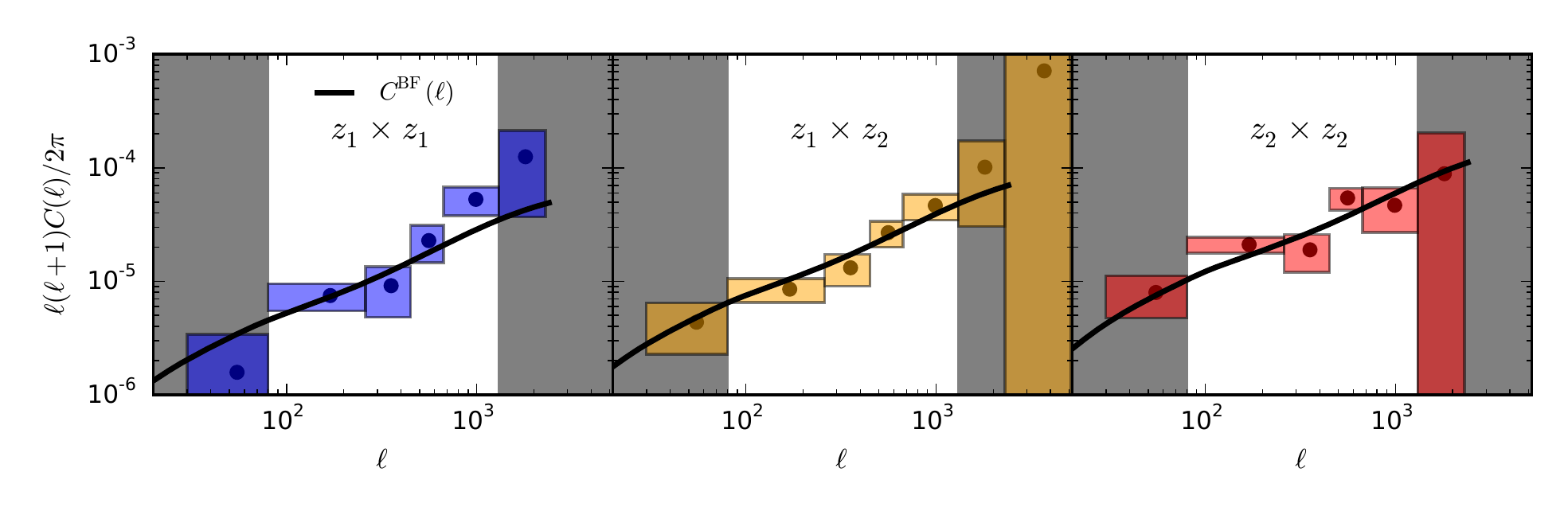}
	\caption{Measured E-mode band powers in tomographic bins averaged with inverse variance weights over all four CFHTLenS patches for illustrative purposes only. From left to right we show the auto-correlation signal of the low redshift bin (blue), the cross-correlation signal between the low and the high redshift bin (orange), and the auto-correlation signal of the high redshift bin (red). The low redshift bin contains objects with redshifts in the range $0.5 < z_1 \leq 0.85$ and the high redshift bin covers a range $0.85<z_2\leq 1.3$. The $1\sigma$-errors in the signal are derived from a run-to-run covariance over 184 independent mock data fields (cf. Section~\ref{sec:covariance}) whereas the extend in $\ell$-direction is the width of the band. Band powers in the shaded regions (grey) to the left and right of each panel are excluded from the cosmological analysis (cf. Fig.~\ref{fig:signals_BB}). The solid line (black) shows the power spectrum for the best fitting, five parameter $\Lambda$CDM model derived in the subsequent analysis (cf. Section~\ref{sec:cosmo_inference} and Table~\ref{tab:results1}). Note, however, that the band powers are centred at the naive $\ell$-bin centre and thus the convolution with the band window function is not taken into account in this plot, in contrast to the cosmological analysis. We present the E-mode signal for each individual CFHTLenS patch in Appendix~\ref{app:figures}.}
	\label{fig:signals_EE}
\end{figure*}

In Fig.~\ref{fig:signals_EE} we show the extracted E-mode band powers for each tomographic bin. For illustrative purposes we combine the band powers extracted from each patch by averaging them with inverse variance weights. The errors on the signal are estimated from the stitched covariance matrix (cf. Section~\ref{sec:covariance}) whereas the extension of the box in $\ell$-direction is just the width of the band. Only bands outside the (grey) shaded areas enter in the cosmological analysis (thus we omit explicitly the `noise catcher' bands, cf. Section~\ref{sec:data_meas} and Table~\ref{tab:bp_intervals}). Note, however, that for the cosmological likelihood analysis we do not use the averaged signals, but instead sum the likelihood of each patch as described in Section~\ref{sec:shear_lkl}.
\begin{figure*}
	\centering
	\includegraphics[width=180mm]{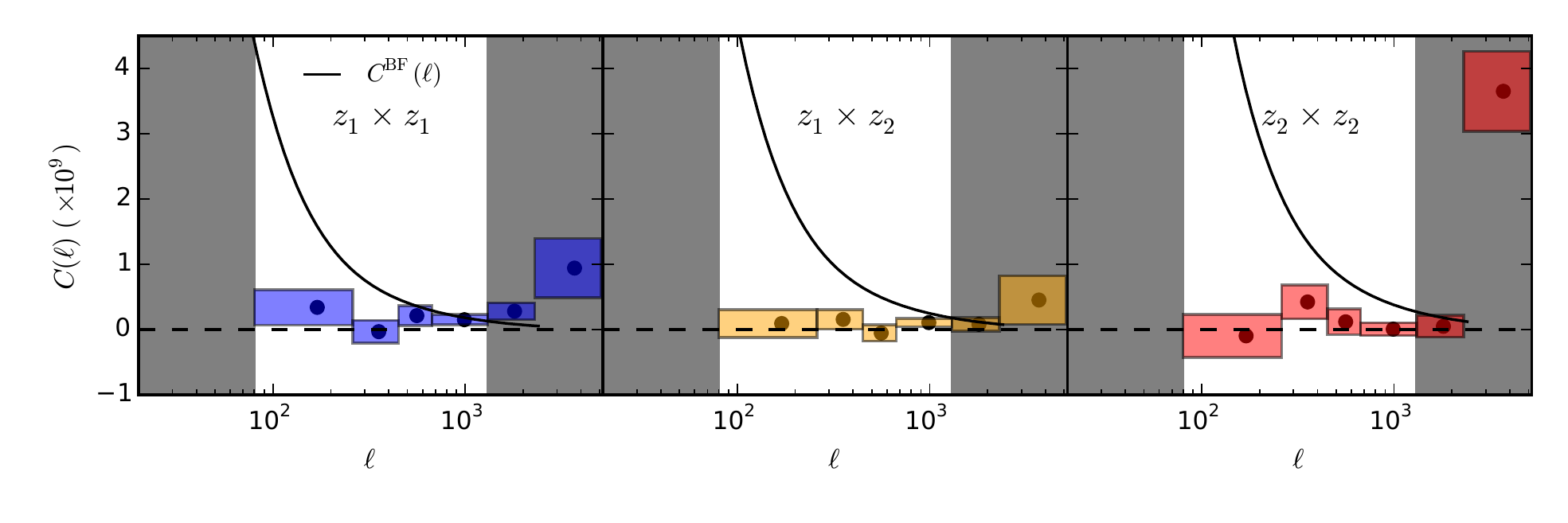}
	\caption{Same as Fig.~\ref{fig:signals_EE} but for B-mode band powers. Note, however, the different scale (linear) and normalization used here with respect to Fig.~\ref{fig:signals_EE}; for reference we also plot the best fitting E-mode power spectrum as solid line (black). We show the measured B-modes as (black) dots with $1\sigma$-errors derived from the inverse Fisher matrix. Based on these signals we define the shaded regions (grey) to the left and right of each panel. E-mode band powers in these regions are excluded from the cosmological analysis (cf. Fig.~\ref{fig:signals_EE} and see text for details). We present the B-mode signal for each individual CFHTLenS patch in Appendix~\ref{app:figures}.}
	\label{fig:signals_BB}
\end{figure*}

We extract E- and B-modes simultaneously. As described in Section~\ref{sec:theory} the cosmological signal is contained in the E-modes in the absence of systematic errors. Hence, we use the B-mode signal as a systematic cross-check and generally expect it to be zero within errors. We do not extract the EB-modes, which would hint at parity-violation in the data, because \citet{Kitching2014} found no evidence for EB-modes in the CFHTLenS data. Hence, we decided to only include the extraction of B-modes as a non-trivial systematic check.
We show the extracted B-mode signal per tomographic bin in Fig.~\ref{fig:signals_BB}. For illustrative purposes we averaged the B-mode signal again with inverse variance weights over all four CFHTLenS patches. In contrast to the E-modes, the $1\sigma$-errors on the B-modes are derived from the B-mode part of the Fisher matrix (cf. equation~\ref{eq:Fisher}). This is a very conservative approach since it will generally underestimate the errorbars. The masking in the data might cause leakage of E-mode power into B-mode power. In principle, this should also be captured by the Fisher matrix but as we argued in Section~\ref{sec:covariance} the Fisher matrix underestimates the E-mode error in the intermediate multipole regime due to the mildly non--Gaussian intrinsic field. This propagates into an underestimated B-mode Fisher-error when compared directly to a run-to-run B-mode error. However, that does not pose a problem as long as we can establish that the B-modes are consistent with zero using the underestimated errorbars. We assess the consistency of the B-modes with zero via a $\chi^2$-goodness-of-fit measure and find: $\chi^2_{\rm red}(W1) = 1.54$, $\chi^2_{\rm red}(W2) = 0.93$, $\chi^2_{\rm red}(W3) = 1.07$, and $\chi^2_{\rm red}(W4) = 0.24$ for 15 degrees of freedom, i.e., including all B-mode bands except the last one, which was designed to catch only noise due to the long tail of the window function of square pixels beyond $\ell_{\rm pix}$. However, further tests conducted on the GRF mock data show that noise from the last band leaks into the second-to-last band depending on the pixel-scale, $\sigma_{\rm pix}$, employed. This is due to the strong oscillatory behaviour of the Fourier-transform of a real-space square pixel (cf. Fig.~2 in \citealt{Hu2001}) around $\ell_{\rm pix}$ corresponding to $\sigma_{\rm pix}$. The oscillations are amplified if the band is noise-dominated. For that reason the B-mode in the second-to-last band appears to be more significant than the B-modes in the other bands. Removing the second-to-last B-mode band power from the $\chi^2$-goodness-of-fit measure yields the following improved reduced $\chi^2$-values for 12 degrees of freedom: $\chi^2_{\rm red}(W1)=0.92$, $\chi^2_{\rm red}(W2)=0.80$, $\chi^2_{\rm red}(W3)=0.61$, and $\chi^2_{\rm red}(W4)=0.23$. Hence, we conclude that the B-modes in these remaining bands are consistent with zero. Therefore, we only use bands 2 to 5 in the cosmological analysis of the E-mode signal.

Following \citet{Becker2015} we define the signal-to-noise ratio, $S/N$, of our band power measurements with respect to the cosmological signal in the mock data from which we estimate the covariance: 
\begin{equation}
S/N = \frac{\bmath{d}_\mathrm{meas}^T \bmath{{\rm C}}_{\mathcal{B}(\ell)}^{-1}\bmath{d}_\mathrm{mock}}{\sqrt{\bmath{d}_\mathrm{mock}^T \bmath{{\rm C}}_{\mathcal{B}(\ell)}^{-1}\bmath{d}_\mathrm{mock}}} \, .
\end{equation}   
Considering only the band powers used in the cosmological analysis (cf. Table~\ref{tab:bp_intervals}), we detect a cosmic shear signal in W1 at $7.1\sigma$, in W2 at $5.5\sigma$, in W3 at $5.7\sigma$, and in W4 only marginally at $2.5\sigma$. Note, however, that the above definition of $S/N$ depends on the cosmology employed in the mocks. A discrepancy between the mock cosmology and the actual cosmology preferred by the data decreases the significance in general. 

\section{Cosmological inference}
\label{sec:cosmo_inference}

After having extracted the shear power spectrum and having derived a more robust estimate of the data covariance, we can proceed to the next step: the cosmological interpretation of the tomographic signals, employing a Bayesian framework. We estimate cosmological parameters $\bmath{p}$ by sampling the likelihood $\mathcal{L}(\bmath{p})$ with a Monte Carlo Markov Chain (MCMC) method. In addition to the parameter estimation we also want to compare various model extensions to a baseline model.

The Bayesian evidence $\mathcal{Z}$ is simply the normalization factor of the posterior over the parameters $\bmath{p}$:
\begin{equation}
\label{eq:evidence}
\mathcal{Z} = \int \diff^{n} \bmath{p} \, \mathcal{L}(\bmath{p}) \pi(\bmath{p}) \, ,
\end{equation}
where $n$ denotes the dimensionality of the parameter space and $\pi(\bmath{p})$ is the prior. Since the evidence is the average of the likelihood over the prior it automatically implements Occam's razor: a simpler theory with fewer parameters, i.e., a more compact parameter space, will have a higher evidence than a more complicated one requiring more parameters, unless the latter model explains the data significantly better. If we wish to decide now between models $M_1$ and $M_0$, we can compare their posterior probabilities given the observed data $\bmath{D}$ and define the Bayes factor:
\begin{equation}
\label{eq:model_selection}
K_{1,0} \equiv \frac{\mathcal{Z}_1}{\mathcal{Z}_0} \frac{\prob(M_1)}{\prob(M_0)} \, ,
\end{equation}  
where $\prob(M_1)/\prob(M_0)$ is the \textit{a priori} probability ratio for the two models, usually set to unity unless there are strong (physical) reasons to prefer one model over the other \textit{a priori}. In our subsequent analysis we always assume $\prob(M_1)/\prob(M_0)=1$. A Bayes factor $K_{1,0} > 1$ implies a preference of model $M_1$ over model $M_0$. \citet{Kass1995} have proposed a quantitative classification scheme for the interpretation of the Bayes factor $K$ (or equivalently $2 \ln K$). 

Evaluating the usually high-dimensional integral of equation~\ref{eq:evidence} is a challenging computational and numerical task. Here, we employ the nested sampling algorithm {\small MULTINEST}\footnote{Version 3.8 from \url{http://ccpforge.cse.rl.ac.uk/gf/project/multinest/}} \citep{Feroz2008, Feroz2009, Feroz2013} via its {\small PYTHON}-wrapper {\small PYMULTINEST} \citep{Buchner2014} in the framework of the cosmological likelihood sampling package {\small MONTE PYTHON}\footnote{Version 2.1.4 from \url{www.montepython.net}} \citep{Audren2012}. 

\subsection{Theoretical power spectrum}
\label{sec:theo_ps}

\begin{figure}
   \centering
	\includegraphics[width=84mm]{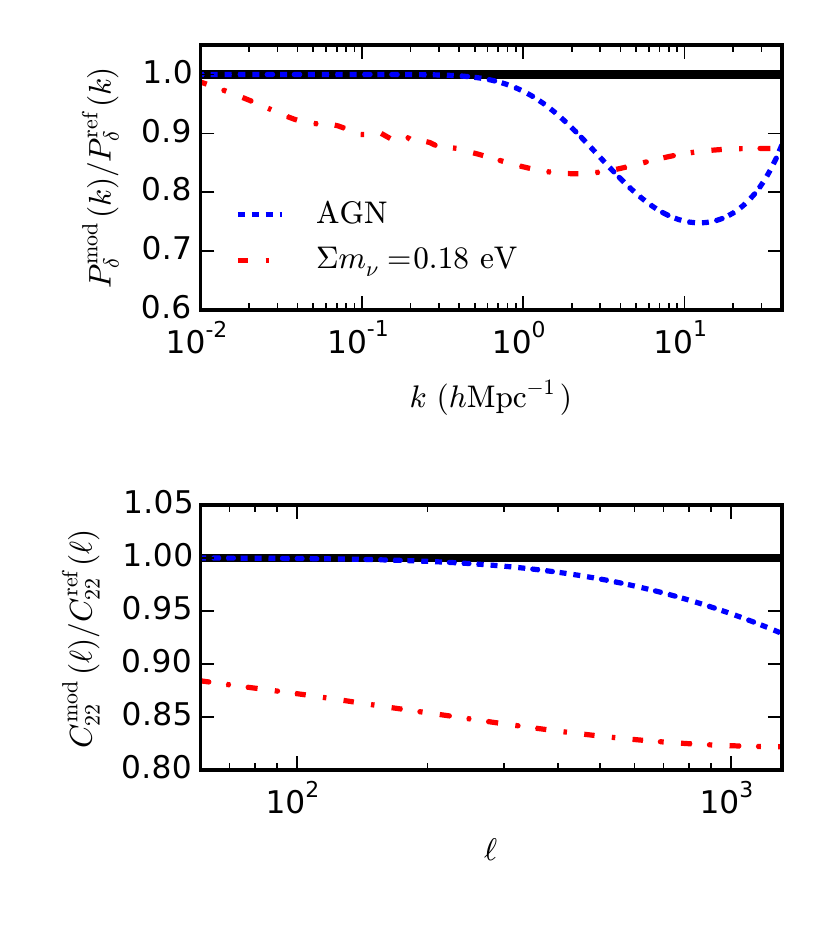}
	\caption{\textbf{Upper panel:} the ratio of modified matter power spectra over the dark matter only power spectrum. The dashed line (blue) shows the effect of the baryon feedback bias in the AGN model from OWLS \citep{Schaye2010, vanDaalen2011} using the implementation by \citet{Harnois2015} (cf. equation~\ref{eq:baryon_feedback}). The modifications due to three degenerate massive neutrinos with total mass $\Sigma m_\nu=0.18 \, {\rm eV}$ is demonstrated by the dash-dotted line (red). The redshift for the power spectrum calculation is $z = 1.05$ corresponding to the median redshift of the high redshift bin used in the subsequent analysis (cf. Table~\ref{tab:n_eff}). \\   
	\textbf{Lower panel:} same as upper panel but for the lensing power spectrum of the high redshift bin $z_2: 0.85 < Z_{\rm B} \leq 1.30$ (cf. Table~\ref{tab:n_eff}).}
	\label{fig:neutrinos_baryons}
\end{figure}

In Section~\ref{sec:theory} we described the calculation of the tomographic lensing power spectra (cf. equation~\ref{eq:theo_power_spec}). These encode the 3D matter power spectrum smoothed by tomographic lensing kernels (cf. equation~\ref{eq:lensing_kernel}). For the calculation of the matter power spectrum, $P_\delta(k; \chi)$, we employ the Boltzmann-code {\small CLASS}\footnote{Version 2.4.3 from \url{www.class-code.net}} \citep{Blas2011, Audren2011}. This already includes the non-linear corrections for which we chose to use the \texttt{halofit} algorithm including the recalibrations by \citet{Takahashi2012}. Furthermore, {\small CLASS} allows us to include (massive) neutrinos \citep{Class_neutrinos}. The main effect of massive neutrinos is a redshift- and scale-dependent reduction of power which also propagates into the lensing power spectra $C_{\ell, \, \mu \nu}^{\mathrm{EE}}$ but is smoothed by the lensing kernels of the corresponding tomographic bins (cf. Fig.~\ref{fig:neutrinos_baryons}). Over the multipole range of interest massive neutrinos lower the lensing power spectrum by an almost constant factor. This introduces a degeneracy with other cosmological parameters that affect the normalization of the lensing power spectrum. 

We follow \citet{Harnois2015} to describe the modifications of the power spectrum due to baryon feedback:
\begin{equation}
b^2(k, z) \equiv \frac{P_{\delta}^{\mathrm{mod}}(k, z)}{P_{\delta}^{\mathrm{ref}}(k, z)} \, ,
\end{equation}
where $P_{\delta}^{\mathrm{mod}}$ and $P_{\delta}^{\mathrm{ref}}$ denote the power spectra with and without baryon feedback, respectively.

The baryon feedback can be computed from hydrodynamical simulations. We use in this work the fitting formula for the baryon feedback derived by \citet{Harnois2015} using the OverWhelmingly Large Simulations (OWLS; \citealt{Schaye2010}, \citealt{vanDaalen2011}):
\begin{equation}
\label{eq:baryon_feedback}
b^2(k, z) = 1 - A_{\mathrm{bary}}(A_z\mathrm{e}^{(B_zx-C_z)^3}-D_zx\mathrm{e}^{E_zx}) \, ,
\end{equation}
where $x=\log_{10}(k/1\mathrm{Mpc}^{-1})$ and the terms $A_z$, $B_z$, $C_z$, $D_z$, and $E_z$ are functions of the scale factor $a = 1/(1+z)$ which are also dependent on the baryonic feedback model (cf. \citealt{Harnois2015} for the specific functional forms and constants). Additionally, we introduce here a general free amplitude $A_{\rm bary}$ which we will use as a free parameter to marginalize over while fitting for the cosmological parameters. In Fig.~\ref{fig:neutrinos_baryons} we show the effect of including baryonic feedback on the matter and lensing power spectrum, respectively. In contrast to the effect of massive neutrinos baryon feedback causes a significant reduction of power in the lensing power spectrum only at high multipoles. However, this is also degenerate with the effect of massive neutrinos on these scales. Hence, a proper anchoring of the main cosmological parameters at low multipoles with high precision is paramount if one wants to break degeneracies between all these effects. Operating directly in multipole space with respect to both theory \textit{and data} facilitates the identification of distinct features in the power spectra. 

\subsection{The shear likelihood}
\label{sec:shear_lkl}

To compare the measured, tomographic band powers $\mathcal{B}^i_\alpha$ (cf. Section~\ref{sec:signal}) to predictions $\langle \mathcal{B}^{i}_\alpha \rangle^{\rm model}$ (cf. Section~\ref{sec:theory}), we define the shear likelihood as a function of cosmological parameters $\bmath{p}$: 
\begin{equation} \label{eq:shear_lkl}
-2 \ln {\mathcal{L}(\bmath{p})} = \sum_i \sum_{\alpha, \, \beta} d_\alpha^i(\bmath{p}) (\bmath{{\rm C}}^{-1})^i_{\alpha \beta} \, d_\beta^i(\bmath{p}) \, ,
\end{equation}
where the index $i$ runs over the four CFHTLenS patches (cf. Section~\ref{sec:data_meas}) and the indices $\alpha$, $\beta$ run over the tomographic bins. Note that we follow all previous CFHTLenS studies in ignoring any covariance between the individual CFHTLenS patches. 

The components of the data vector per patch are calculated as
\begin{equation}
d_\alpha^i(\bmath{p}) = (\mathcal{B}^i_\alpha - \langle \mathcal{B}^{i}_\alpha(\bmath{p}) \rangle^{\rm model}) \, ,
\end{equation}
where the predicted band powers, $\langle \mathcal{B}^{i}(\ell) \rangle^{\rm model}$, depend on the cosmological parameters $\bmath{p}$. They are calculated with equations~\ref{eq:conv_window_func}, \ref{eq:theo_power_spec}, i.e., the band window functions are properly taken into account.
The inverse of the covariance matrix $\bmath{{\rm C}}^{-1}$ is estimated from a large suite of mock data especially tailored to CFHTLenS as described in detail in Section~\ref{sec:covariance}.

\subsection{Models \& discussion}
\label{sec:models_discussion}

In the first part of this cosmological analysis we consider the shear likelihood without further combining it with any other external cosmological probe. The lensing power spectrum is most sensitive to cosmological parameters modifying its normalization and slope. Therefore, the normalization of the primordial power spectrum, $\ln(10^{10}A_s)$, and the fraction of cold dark matter, $\Omega_{\rm cdm}$, are the primary parameters of interest. For an easier comparison of our results with the literature, we also derive the root-mean-square variance of the density field smoothed with the Fourier transform of a top-hat filter on a scale ${\rm R}=8 \, h^{-1} {\rm Mpc}$ in real-space, i.e. $\sigma_8$, and the total fraction of matter in the Universe, $\Omega_{\rm m}$. 
Our baseline model to which we refer subsequently only as `$\Lambda$CDM' includes in addition to these parameters three more free variables: the Hubble parameter $h$, the slope of the primordial power spectrum $n_{\rm s}$, and the fraction of baryonic matter $\Omega_{\rm b}$. The ranges for the flat priors on these parameters are listed in Table~\ref{tab:results1}. They follow mostly the ranges employed in the CFHTLenS studies by \citet{Benjamin2013} and \citet{Heymans2013} in order to assure a fair comparison of our results with these studies.  

Standard model neutrinos have mass (e.g. \citealt{Lesgourgues2006} and references therein). Hence, we follow \citet{Planck2015_CP} in including already two massless and one massive neutrino with the (fixed) minimal mass of $\Sigma m_\nu = 0.06 \, {\rm eV}$ (assuming a normal mass hierarchy with one dominant mass eigenstate) in our baseline $\Lambda$CDM model. Moreover, we always assume a flat cosmological model. %Note that we decided to lower the upper bound of the prior for the Hubble parameter, $h$, from 1.2 to 1. 

The first extension of the baseline model is to introduce a free total mass $\Sigma m_\nu$ for three degenerate massive neutrinos. We refer to this model as `$\Lambda$CDM$+\nu$'. Since we expect the effect of massive neutrinos to be degenerate with the effect of baryonic feedback, especially at high multipoles (cf. Section~\ref{sec:theory} and Fig.~\ref{fig:neutrinos_baryons}) we investigate this effect in the model `$\Lambda$CDMa': here, we additionally include the fiducial baryon feedback model of equation~\ref{eq:baryon_feedback} with $A_{\mathrm{bary}}=1$ for the AGN model taken from the OWLS project \citep{Schaye2010, vanDaalen2011}. 
The degeneracy between baryonic feedback and massive neutrinos is investigated in the model $\Lambda$CDMa$+\nu$, where $\Sigma m_\nu$ is free to vary but which includes the fixed fiducial baryon feedback model.
We relax the assumption of a fixed baryon feedback model in the model `$\Lambda$CDM$+A_{\mathrm{bary}}$' by allowing the amplitude of the feedback $A_{\mathrm{bary}}$ to vary (cf. equation~\ref{eq:baryon_feedback}).
Combining the assumption of a free amplitude in the baryon feedback model and a free total mass of three degenerate massive neutrinos, $\Sigma m_\nu$ in the model `$\Lambda \mathrm{CDM}+\nu+A_\mathrm{bary}$' yields a maximally degenerate model in baryonic feedback and neutrinos. In total this model consists of seven free parameters. 

Moreover, we want to test the effect of a photometric redshift bias which causes a coherent shift of the photometric redshift distributions per tomographic bin (cf. equation~\ref{eq:lensing_kernel}) by $\Delta z_\mu$. \citet{Hildebrandt2012} showed that the bias on photometric redshifts in CFHTLenS is $\Delta z < 0.02$ (cf. their Fig.~8). However, this estimate does not account for outliers which can increase the photometric redshift bias significantly. Therefore, we make a more conservative assumption and treat the photometric redshift biases $\Delta z_\mu$ within a flat prior range of $-0.05 \leq \Delta z_\mu \leq 0.05$ as nuisance parameters to marginalize over.    
In the most complex model $\Lambda \mathrm{CDM}+\nu+A_\mathrm{bary}+\Delta z_\mu$, which we abbreviate subsequently to `$\Lambda{\rm CDM}+{\rm all}$', we include a free amplitude for the baryon feedback model, massive neutrinos and treat the photometric redshift biases $\Delta z_\mu$ as nuisance parameters.

All models, their prior ranges and the parameter estimates derived from the likelihood sampling are summarized in Table~\ref{tab:results1}, where we always quote the weighted median value for each varied parameter. The errors denote the 68 per cent credible interval of the posterior distribution after marginalization over all other free parameters.
\begin{table*}
	\caption{Cosmological parameters from shear likelihood only}
	\label{tab:results1}
	\begin{center}
		\resizebox{\textwidth}{!}{%
		\begin{tabular}{ c c c c c c c c c c c c }
			\hline
			Model& $\Omega_\mathrm{cdm}$& $\ln(10^{10}A_\mathrm{s})$& $\Omega_\mathrm{m}$& $\sigma_8$& $\Omega_\mathrm{b}$& $n_\mathrm{s}$& $h$& $\Sigma m_\nu \, ({\rm eV})$& $A_\mathrm{bary}$& $\Delta z_1$& $\Delta z_2$\\
			\hline
			Prior ranges& $[0., 1.]$& $[0., 10.]$& derived& derived& $[0., 0.1]$& $[0.7, 1.3]$& $[0.4, 1.]$& $[0.06, 6.]$& $[0., 10.]$& $[-0.05, 0.05]$& $[-0.05, 0.05]$\\
			\hline			
			$\Lambda \mathrm{CDM}$& $0.21_{-0.15}^{+0.09}$& $3.53_{-1.52}^{+1.49}$& $0.26_{-0.15}^{+0.09}$& $0.84_{-0.23}^{+0.24}$& $0.05_{-0.03}^{+0.03}$& $1.01_{-0.23}^{+0.29}$&  	$0.62_{-0.22}^{+0.09}$& $\equiv 0.06$& --& --& --\\
			$\Lambda \mathrm{CDMa}$& $0.21_{-0.14}^{+0.09}$& $3.50_{-1.62}^{+1.43}$& $0.25_{-0.15}^{+0.11}$& $0.85_{-0.24}^{+0.24}$& $0.05_{-0.03}^{+0.02}$& $1.00_{-0.22}^{+0.26}$&  	$0.64_{-0.22}^{+0.10}$& $\equiv 0.06$& $\equiv 1.$& --& --\\
			$\Lambda \mathrm{CDM}+\nu$& $0.21_{-0.13}^{+0.08}$& $3.65_{-1.44}^{+1.52}$& $0.30_{-0.14}^{+0.09}$& $0.75_{-0.15}^{+0.16}$& $0.04_{-0.03}^{+0.02}$& $1.05_{-0.28}^{+0.25}$&  	$0.70_{-0.16}^{+0.18}$& $1.37_{-1.31}^{+0.69}$& --& --& --\\
			%$M_{1b}$& & & & & & & & \\
			$\Lambda \mathrm{CDMa}+\nu$& $0.21_{-0.13}^{+0.09}$& $3.69_{-1.52}^{+1.37}$& $0.29_{-0.14}^{+0.11}$& $0.76_{-0.16}^{+0.16}$& $0.04_{-0.03}^{+0.02}$& $1.05_{-0.27}^{+0.25}$&  	$0.71_{-0.18}^{+0.22}$& $1.34_{-1.28}^{+0.60}$& $\equiv 1.$& --& --\\			
			$\Lambda \mathrm{CDM}+A_\mathrm{bary}$& $0.21_{-0.14}^{+0.10}$& $3.62_{-1.47}^{+1.51}$& $0.26_{-0.14}^{+0.10}$& $0.85_{-0.26}^{+0.25}$& $0.05_{-0.03}^{+0.02}$& $1.00_{-0.24}^{+0.19}$&  	$0.60_{-0.20}^{+0.09}$& $\equiv 0.06$& $2.90_{-2.90}^{+1.54}$& --& --\\
			$\Lambda \mathrm{CDM}+\nu+A_\mathrm{bary}$& $0.22_{-0.13}^{+0.08}$& $3.69_{-1.42}^{+1.44}$& $0.30_{-0.15}^{+0.09}$& $0.76_{-0.15}^{+0.15}$& $0.04_{-0.03}^{+0.02}$& $1.06_{-0.28}^{+0.24}$&  	$0.69_{-0.17}^{+0.17}$& $1.29_{-1.23}^{+0.67}$& $2.51_{-2.51}^{+1.19}$& --& --\\
			$\Lambda \mathrm{CDM}+\Delta z_\mu$& $0.24_{-0.14}^{+0.10}$& $3.26_{-1.32}^{+1.28}$& $0.29_{-0.15}^{+0.10}$& $0.80_{-0.22}^{+0.21}$& $0.05_{-0.03}^{+0.03}$& $0.98_{-0.21}^{+0.19}$&  	$0.62_{-0.21}^{+0.10}$& $\equiv 0.06$& --& $0.03_{-0.01}^{+0.02}$& $-0.02_{-0.03}^{+0.02}$\\
			$\Lambda \mathrm{CDM}+{\rm all}$& $0.24_{-0.13}^{+0.09}$& $3.57_{-1.44}^{+1.34}$& $0.32_{-0.13}^{+0.10}$& $0.74_{-0.14}^{+0.14}$& $0.04_{-0.03}^{+0.02}$& $1.04_{-0.25}^{+0.26}$&  	$0.67_{-0.17}^{+0.16}$& $1.32_{-1.26}^{+0.56}$& $2.49_{-2.49}^{+1.17}$& $0.03_{-0.01}^{+0.02}$& $-0.02_{-0.03}^{+0.02}$\\
			\hline
		\end{tabular}}
	\end{center}
	\medskip 
	\textit{Notes.} We quote weighted median values for each varied parameter and derive $1\sigma$-errors using the 68 per cent credible interval of the marginalized posterior distribution.
\end{table*}

\begin{figure}
	\centering
	\includegraphics[width=84mm]{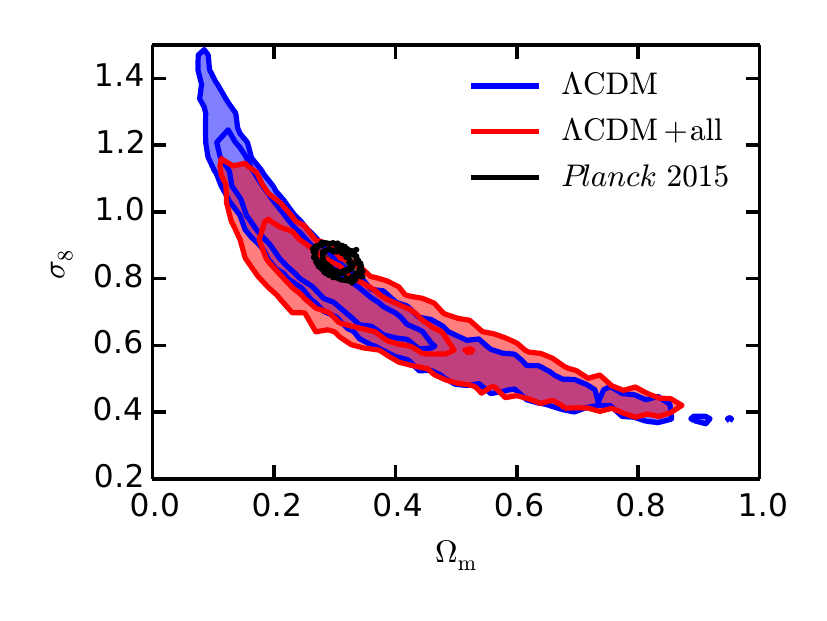}
	\caption{Shown are $68$ and $95$ per cent credible intervals (blue, inner and outer contour, respectively) for our baseline $\Lambda$CDM model where $\Omega_m$, $\sigma_8$, $h$, $n_s$, and $\Omega_b$ are free to vary. Additionally shown are the $68$ and $95$ per cent credible intervals (red,  respectively) for our most complex model $\Lambda \mathrm{CDM}+{\rm all}$ where also the total mass of neutrinos $\Sigma m_\nu$, the amplitude for the Baryon feedback model $A_\mathrm{bary}$, and a systematic photometric redshift bias per tomographic bin $\Delta z_\mu$ are free to vary. We marginalize over all other free parameters. Finally, we plot the $68$ and $95$ per cent credible intervals derived from \citet{Planck2015_CP} (TT+lowP).}
	\label{fig:bl_omega_sigma_plane}
\end{figure}

\begin{figure*}
	\centering
	\includegraphics[width=180mm]{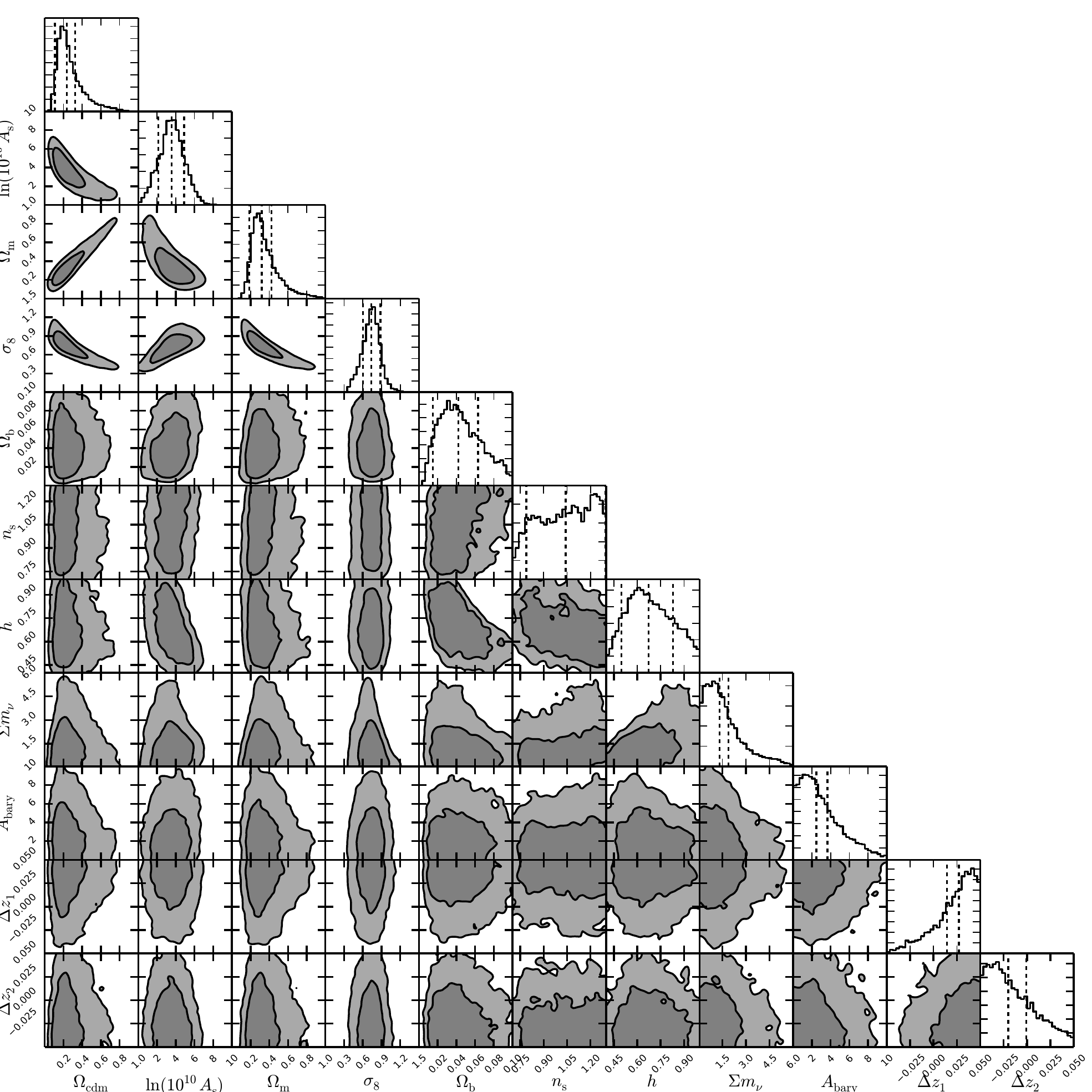}
	\caption{Shown are all parameter constraints from sampling the likelihood of model $\Lambda {\rm CDM}+{\rm all}$. The dashed lines in the marginalized 1D posteriors denote the weighted median and the 68 per cent credible interval (cf. Table~\ref{tab:results1}). The contours in each 2D likelihood contour subplot are $68$ and $95$ per cent credible intervals smoothed with a Gaussian.}
	\label{fig:triangle_base_ext_all}
\end{figure*}

We compare the 68 and 95 per cent credible intervals for the baseline $\Lambda$CDM model and the most complex $\Lambda \mathrm{CDM}+{\rm all}$ model in Fig.~\ref{fig:bl_omega_sigma_plane}. Both models are marginally consistent with the 68 per cent credible interval from \citet{Planck2015_CP} (TT+lowP) and the most complex $\Lambda \mathrm{CDM}+{\rm all}$ model is fully consistent with \textit{Planck} at 95 per cent credibility. 
This model is very conservative because it also accounts for a possible photometric redshift bias per tomographic bin, and thus is expected to yield the largest errorbars. For this model we show marginalized 1D posteriors for every free parameter (cf. Table~\ref{tab:results1}) and marginalized 2D contours for every parameter combination in Fig.~\ref{fig:triangle_base_ext_all}. From this figure but also from Table~\ref{tab:results1} it is apparent that our parameter constraints are weaker than those derived from \textit{Planck}. The shear data are also unable to constrain the slope of the primordial power spectrum, $n_{\rm s}$, especially once the models also include massive neutrinos, since both parameters influence the slope of the lensing power spectrum in a similar way. Hence, the estimate on $n_{\rm s}$ is following the flat prior distribution.   
From our most conservative model extension, $\Lambda$CDM+all, we derive an upper limit on the total mass of three degenerate massive neutrinos at 95 per cent credibility of $\Sigma m_\nu < 4.53 \, {\rm eV}$. In contrast, \citet{Planck2015_CP} (TT+lowP) derive an upper limit (95 per cent) on the total mass of three degenerate massive neutrinos of $\Sigma m_\nu < 0.72 \, {\rm eV}$. Combining the primary CMB data with secondary data and/or other external probes lowers the upper limit to $<0.17 \, {\rm eV}$. 

\begin{figure}
	\centering
	\includegraphics[width=84mm]{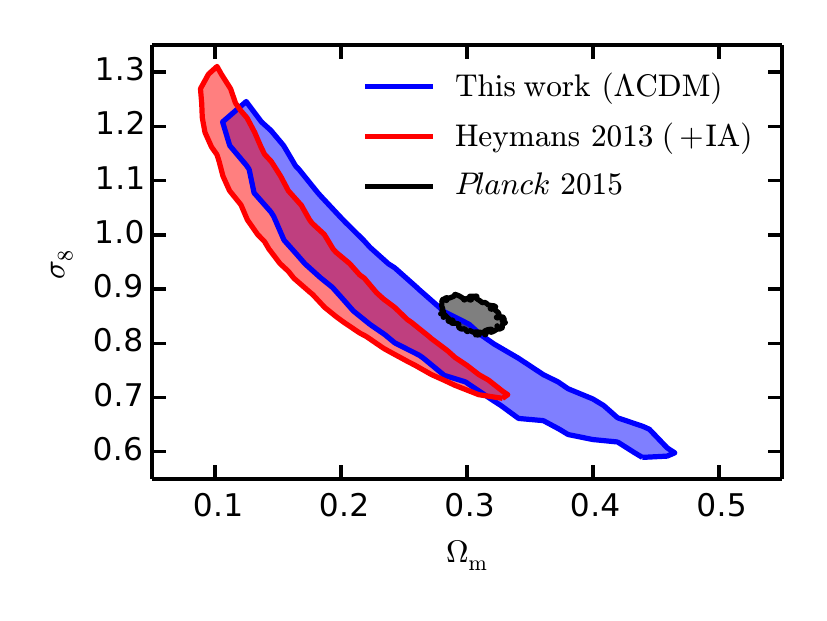}
	\caption{We show the $68$ per cent credible interval (blue) for our baseline $\Lambda$CDM model. Additionally shown is the $68$ per cent credible interval for the 6-bin tomographic real space analysis from \citet{Heymans2013} (cf. also their figure 4) where intrinsic alignments are marginalized over (red). Finally, we plot the $68$ per cent credible interval from \citet{Planck2015_CP} (TT+lowP).}
	\label{fig:comp_contour_heymans2013}
\end{figure}

In the $\sigma_8$--$\Omega_{\rm m}$-plane we can directly compare to the results from the CFHTLenS analysis by \citet{Heymans2013}. They employed a 6-bin tomographic real-space correlation approach and in Fig.~\ref{fig:comp_contour_heymans2013} we show the 68 per cent credible intervals for their conservative model including a marginalization over intrinsic alignments. The $68$ per cent credible intervals of our baseline $\Lambda$CDM model is consistent with the one derived by \citet{Heymans2013}. However, the contours of our model are generally broader because we use only two tomographic bins.

\begin{figure}
	\centering
	\includegraphics[width=84mm]{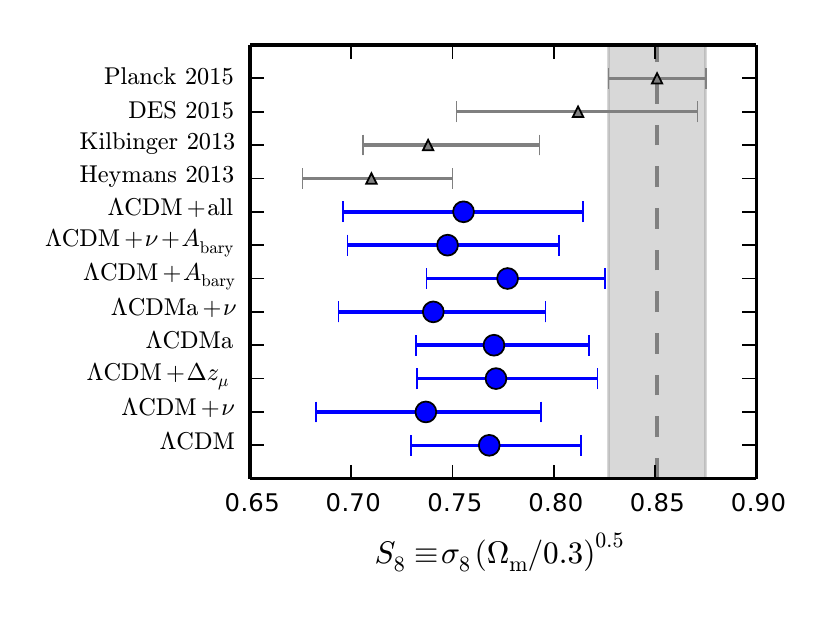}
	\caption{Shown are $1\sigma$-constraints on the parameter combination $S_8 \equiv \sigma_8(\Omega_\mathrm{m}/0.3)^{0.5}$ for all of our tested models (cf. Tables~\ref{tab:results1} and ~\ref{tab:combination}). We compare them to constraints from other lensing analyses and to the constraint from \citet{Planck2015_CP} (TT+lowP). Note that for \citet{Heymans2013} and \citet{Kilbinger2013} we quote the values derived in \citet{DES_cosmo2015} for the ``original conservative scales'' and for ``all scales'', respectively. `DES 2015' refers to the fiducial result from \citet{DES_cosmo2015} (``Fiducial DES SV cosmic shear'').}
	\label{fig:Kilbinger-style_S8}
\end{figure}

The shear power spectrum is most sensitive to the parameters $\Omega_\mathrm{cdm}$ and $\ln(10^{10}A_\mathrm{s})$ or equivalently to $\Omega_\mathrm{m}$ and $\sigma_8$. However, as can be seen in, for example, Fig.~\ref{fig:bl_omega_sigma_plane} the relation between $\Omega_\mathrm{m}$ and $\sigma_8$ is degenerate and what lensing can actually constrain best is the combination of both parameters in the projected quantity $\sigma_8(\Omega_\mathrm{m}/0.3)^\alpha$. The value of $\alpha$ depends on the scales probed and is connected to the width of the likelihood contour. We derive it from fitting the function $\ln \sigma_8 (\Omega_{\rm m}) = -\alpha \ln \Omega_{\rm m}+{\rm const.}$ to the likelihood surface in the $\sigma_8$--$\Omega_{\rm m}$-plane. Since we find it to be consistent with $\approx 0.5$ in all our models, we follow \citet{DES_cosmo2015} in defining the quantity $S_8 \equiv \sigma_8(\Omega_\mathrm{m}/0.3)^{0.5}$. We present values for this parameter combination obtained from the above shear-only likelihood sampling in Table~\ref{tab:combination}. We compare the values of $S_8$ for all our models in Fig.~\ref{fig:Kilbinger-style_S8}, where we also show the constraint on that parameter combination by \citet{Planck2015_CP} (TT+lowP). For this combination all our tested models are consistent with each other. However, all models are in mild tension with the constraint on $S_8$ derived from \textit{Planck} (TT+lowP).
\begin{table*}
	\begin{minipage}{180mm}
	\caption{Constraints on $S_8$ and $\sigma_8(\Omega_\mathrm{m}/0.3)^\alpha$}
	\label{tab:combination}
	\begin{center}
		%\resizebox{\columnwidth}{!}{%
		\begin{tabular}{ c c c c c }
			\hline
			Model& $S_8 \equiv \sigma_8(\Omega_\mathrm{m}/0.3)^{0.5}$& Mean error on $S_8$&$\sigma_8(\Omega_\mathrm{m}/0.3)^\alpha$& $\alpha$\\
			\hline
			\textit{Shear likelihood only}\\
			$\Lambda \mathrm{CDM}$& $0.768_{-0.039}^{+0.045}$& 0.042& $0.762_{-0.038}^{+0.044}$& 0.538\\
			$\Lambda \mathrm{CDMa}$& $0.770_{-0.039}^{+0.047}$& 0.043& $0.765_{-0.038}^{+0.044}$& 0.533\\		
			$\Lambda \mathrm{CDM}+\nu$& $0.737_{-0.054}^{+0.057}$& 0.056& $0.737_{-0.055}^{+0.057}$& 0.479\\
			$\Lambda \mathrm{CDMa}+\nu$& $0.741_{-0.047}^{+0.055}$& 0.051& $0.741_{-0.046}^{+0.056}$& 0.465\\
			$\Lambda \mathrm{CDM}+A_\mathrm{bary}$& $0.777_{-0.040}^{+0.048}$& 0.044& $0.773_{-0.040}^{+0.046}$& 0.531\\
			$\Lambda \mathrm{CDM}+\nu+A_\mathrm{bary}$& $0.748_{-0.049}^{+0.055}$& 0.052& $0.748_{-0.050}^{+0.054}$& 0.479\\
			$\Lambda \mathrm{CDM}+\Delta z_\mu$& $0.771_{-0.039}^{+0.050}$& 0.045& $0.767_{-0.037}^{+0.045}$& 0.555\\
			$\Lambda \mathrm{CDM}+{\rm all}$& $0.755_{-0.059}^{+0.059}$& 0.059& $0.755_{-0.059}^{+0.059}$& 0.491\\
			\hline
		\end{tabular}%}
	\end{center}
	\medskip 
	\textit{Notes.} We quote median values for the constraints on $S_8 \equiv \sigma_8(\Omega_\mathrm{m}/0.3)^{0.5}$ and $\sigma_8(\Omega_\mathrm{m}/0.3)^\alpha$. The errors denote the 68 per cent credible interval derived from the marginalized posterior distribution.
	\end{minipage}
\end{table*}

Moreover, we present in Fig.~\ref{fig:Kilbinger-style_S8} the constraints on $S_8$ of other lensing studies. In particular, we compare to the recent constraint from \citet{DES_cosmo2015} (``Fiducial DES SV cosmic shear''). This study employed a real-space correlation function approach in three tomographic bins. We find our constraints to be consistent with theirs which is mainly due to the large errorbars of the measurement on $S_8$ from the Dark Energy Survey (DES). In addition to their own results \citet{DES_cosmo2015} also resampled the likelihoods of the CFHTLenS studies from \citet{Kilbinger2013} and \citet{Heymans2013} and derived constraints on $S_8$. We show these constraints also in Fig.~\ref{fig:Kilbinger-style_S8}. \citet{Kilbinger2013} employed a non-tomographic real-space correlation function approach and their constraint in Fig.~\ref{fig:Kilbinger-style_S8} employs ``all scales'' out to large angular scales $\vartheta \approx 350 \arcmin$. The constraint from \citet{Heymans2013} in Fig.~\ref{fig:Kilbinger-style_S8} uses only the ``original conservative scales''. Our results are consistent with both these studies, as was already the case for \citet{Heymans2013} in the full $\sigma_8$--$\Omega_{\rm m}$-plane (cf. Fig.~\ref{fig:comp_contour_heymans2013}). % and additionally our baseline $\Lambda$CDM model is marginally consistent with the constraint from \textit{Planck}, as was already the case in the full $\sigma_8$--$\Omega_{\rm m}$-plane (cf. Fig.~\ref{fig:bl_omega_sigma_plane}).

\begin{figure}
	\centering
	\includegraphics[width=84mm]{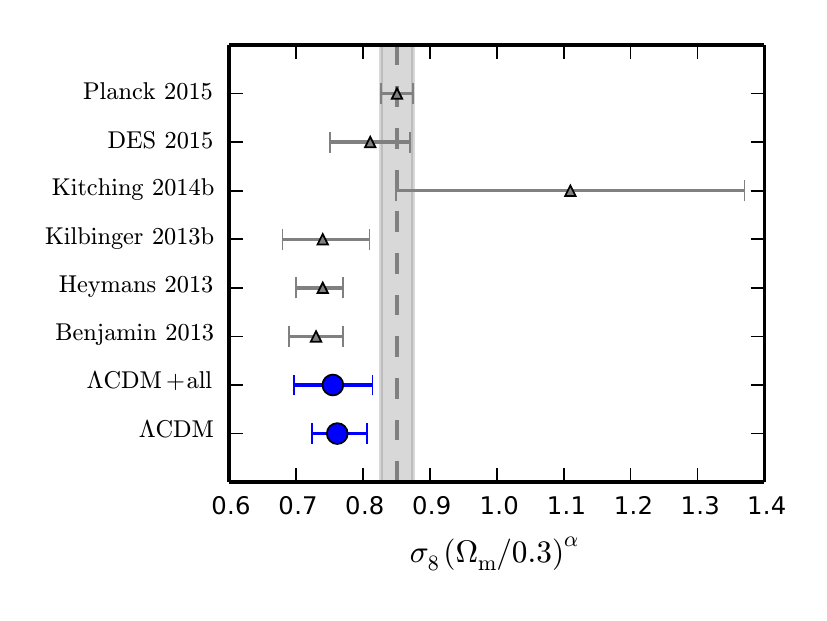}
	\caption{Shown are $1\sigma$-constraints on the parameter combination $\sigma_8(\Omega_\mathrm{m}/0.3)^\alpha$ for our $\Lambda$CDM and $\Lambda{\rm CDM}+{\rm all}$ models (cf. Table~\ref{tab:results1}). We compare them to constraints from other lensing analyses and to the constraint from \citet{Planck2015_CP} (TT+lowP). Note that for \citet{Heymans2013} we quote the value derived by marginalising over intrinsic alignments. For \citet{Kilbinger2013} and \citet{Kitching2014} we cite values including the largest scales in their analyses.}
	\label{fig:Kilbinger-style_comp_ext_models}
\end{figure}

For the comparison of our results to other CFHTLenS studies and the originally published constraints from \citet{Kilbinger2013} and \citet{Heymans2013} we have to resort to the parameter combination $\sigma_8(\Omega_\mathrm{m}/0.3)^\alpha$. The exponent $\alpha$ is in general quite similar between the quoted lensing studies but not the same, which the reader should bear in mind when looking at Fig.~\ref{fig:Kilbinger-style_comp_ext_models}. For completeness, we show again the constraints on that parameter combination from \citet{Heymans2013} and \citet{Kilbinger2013}. \citet{Kitching2014} employed a 3D lensing approach which allows for control over the $k$-scales included in the analysis. However, their constraint on $\sigma_8(\Omega_\mathrm{m}/0.3)^\alpha$ for which we quote the value including large scales, i.e., $k \leq 5 \, h {\rm Mpc}^{-1}$, yields by far the largest errorbars due to which their constraint is consistent with all other CFHTLenS studies and also consistent with the \textit{Planck} constraint. The analysis by \citet{Benjamin2013} is the most similar to the one presented here: although their analysis employed a real-space correlation function approach and did not include scales as large as the ones used here, the two redshift bins in their tomographic analysis are exactly the same ones employed in this analysis. The constraints are also consistent with each other and especially our $\Lambda$CDM model also yields comparable errorbars.

In summary, all models are consistent with each other mainly due to increasing errorbars for increasingly more free parameters. For the comparison of our analysis to other cosmic shear studies we derived a constraint on the projected parameters $S_8$ or $\sigma_8(\Omega_\mathrm{m}/0.3)^\alpha$. In general, we find consistency in these projected parameters with all other CFHTLenS studies and DES. 
Employing the Bayesian model comparison framework, we can decide which of the tested models describes the shear data best: in Table~\ref{tab:evidences} we present the natural logarithms of the evidence for each model. Comparing these models in terms of their Bayes factor $K$ with respect to the simplest models $\Lambda$CDM or $\Lambda$CDMa, we find no evidence for any of the tested extensions except for a very weak preference of the model $\Lambda$CDM$+\Delta z_\mu$ over our baseline model which is according to the interpretation scheme of \citet{Kass1995} ``not worth more than a bare mention''.
\begin{table*}
	\begin{minipage}{180mm}
	\caption{Evidences from shear likelihood only}
	\label{tab:evidences}
	\begin{center}
		\resizebox{\textwidth}{!}{%
		\begin{tabular}{ c c c c }
			\hline
			Model& $\ln \mathcal{Z}$& $2 \ln K$ ($K\equiv\mathcal{Z}_i/\mathcal{Z}_{\Lambda \mathrm{CDM}}$)& interpretation\\
			\hline
			$\Lambda \mathrm{CDM}$& $-40.96 \pm 0.06$& 0& --\\
			$\Lambda \mathrm{CDMa}$& $-41.07 \pm 0.06$& -0.22& support for $\Lambda$CDM\\		
			$\Lambda \mathrm{CDM}+\nu$& $-41.63 \pm 0.07$& -1.34& support for $\Lambda$CDM\\
			$\Lambda \mathrm{CDMa}+\nu$& $-41.83 \pm 0.07$& -1.74& support for $\Lambda$CDM\\
			$\Lambda \mathrm{CDM}+A_\mathrm{bary}$& $-41.66 \pm 0.06$& -1.40& support for $\Lambda$CDM\\
			$\Lambda \mathrm{CDM}+\nu+A_\mathrm{bary}$& $-42.48 \pm 0.07$& -3.04& support for $\Lambda$CDM\\
			$\Lambda \mathrm{CDM}+\Delta z_\mu$& $-40.75 \pm 0.07$& 0.42& preference over $\Lambda$CDM ``not worth more than a bare mention''\\
			$\Lambda \mathrm{CDM}+{\rm all}$& $-42.19 \pm 0.07$& -2.46& support for $\Lambda$CDM\\
			\hline
		\end{tabular}}
	\end{center}
	\medskip 
	\textit{Notes.} For each model we calculate the global log-evidence, $\ln \mathcal{Z}$, and compare all evidences in terms of the Bayes factor $K$ (or equivalently $2\ln K$) to the baseline $\Lambda$CDM model. The interpretation of the Bayes factor is following the scheme proposed by \citet{Kass1995}.
	\end{minipage}
\end{table*}

Hence, we conclude that the extracted band powers of the tomographic shear power spectra measured over a range $80 \leq \ell \leq 1310$ are described sufficiently well within their errors by a standard, five parameter $\Lambda$CDM model.

Finally, we combine our shear likelihood with the most recent data and likelihood release\footnote{PLC-2.0 from \url{http://pla.esac.esa.int/pla/}} from \citet{Planck2015_CP} in order to break the degeneracy between the parameters $\Omega_{\rm m}$ and $\sigma_8$. In particular we employ the \textit{Planck} primary CMB temperature data (TT) from high multipoles $\ell$ in combination with the \textit{Planck} low multipole polarization data (lowP). Due to long run-time we chose to use the {\small PLIK HIGHL-LITE} likelihood code which requires only to marginalize over one nuisance parameter, $A_\mathrm{Planck}$. The Bayesian model comparison showed no evidence for any model extension beyond a baseline $\Lambda$CDM model for describing the shear likelihood. This implies that we would essentially reproduce \textit{Planck}-only results if we were to add parameters for which there is no evidence. Hence, we consider only six cosmological parameters and one nuisance parameter for the combined  `\textit{Planck}+Shear' model: $\Omega_\mathrm{cdm}$, $\ln(10^{10}A_\mathrm{s})$, $h$, $\Omega_{\rm b}$, $n_{\rm s}$, $\tau_{\rm reio}$, and $A_{\rm Planck}$. Again we assume one dominant neutrino mass eigenstate in the normal hierarchy with $\Sigma m_\nu = 0.06 \, {\rm eV}$ and a flat cosmology. In comparison to our shear-only likelihood analysis we chose to use narrower prior ranges for most parameters (cf. Table~\ref{tab:results2}). Due to the reduced set of nuisance parameters with respect to the original \textit{Planck} analysis, we also resample the \textit{Planck} likelihood for the seven parameter baseline model so that comparisons of likelihood contours are fair.
\begin{table*}
	\caption{Cosmological parameters from a combined analysis of the shear and \textit{Planck} likelihoods}
	\label{tab:results2}
	\begin{center}
		\resizebox{\textwidth}{!}{%
		\begin{tabular}{ c c c c c c c c c c }% c }
			\hline
			Model& $\Omega_\mathrm{cdm}$& $\ln(10^{10}A_\mathrm{s})$& $\Omega_\mathrm{m}$& $\sigma_8$& $\Omega_\mathrm{b}$& $n_\mathrm{s}$& $h$& $\tau_{\rm reio}$& $A_{\rm Planck}$\\
			\hline
			Prior ranges& $[0.1, 0.4]$& $[2., 4.]$& derived& derived& $[0., 0.1]$& $[0.8, 1.2]$& $[0.5, 0.8]$& $[0.04, 0.12]$& $[90., 110.]$\\
			\hline
			\textit{Planck} (TT+lowP)& $0.263_{-0.013}^{+0.012}$& $3.093_{-0.034}^{+0.037}$& $0.313_{-0.014}^{+0.013}$& $0.830_{-0.015}^{+0.014}$& $0.049_{-0.001}^{+0.001}$& $0.966_{-0.006}^{+0.007}$&  	$0.674_{-0.010}^{+0.010}$& $0.079_{-0.019}^{+0.018}$& $100.04_{-0.26}^{+0.27}$\\
			%old data vector:
			%\textit{Planck}+Shear& $0.249_{-0.010}^{+0.010}$& $3.078_{-0.036}^{+0.038}$& $0.298_{-0.011}^{+0.011}$& $0.817_{-0.014}^{+0.013}$& $0.048_{-0.001}^{+0.001}$& $0.971_{-0.006}^{+0.006}$&  	$0.685_{-0.009}^{+0.008}$& $0.074_{-0.019}^{+0.019}$& $100.03_{-0.27}^{+0.27}$\\
			%new data vector:			
			\textit{Planck}+Shear& $0.251_{-0.010}^{+0.010}$& $3.077_{-0.035}^{+0.037}$& $0.300_{-0.011}^{+0.011}$& $0.818_{-0.013}^{+0.013}$& $0.048_{-0.001}^{+0.001}$& $0.971_{-0.006}^{+0.006}$&  	$0.684_{-0.009}^{+0.008}$& $0.074_{-0.018}^{+0.020}$& $100.02_{-0.26}^{+0.27}$\\
			\hline
		\end{tabular}}
	\end{center}
	\medskip 
	\textit{Notes.} We quote weighted median values for each varied parameter and derive $1\sigma$-errors using the 68 per cent credible interval of the marginalized posterior distribution. For the model \textit{Planck} (TT+lowP) we resampled a simplified version of the original likelihood that includes only one additional nuisance parameter, $A_{\rm Planck}$.
\end{table*} 

\begin{figure}
	\centering
	\includegraphics[width=84mm]{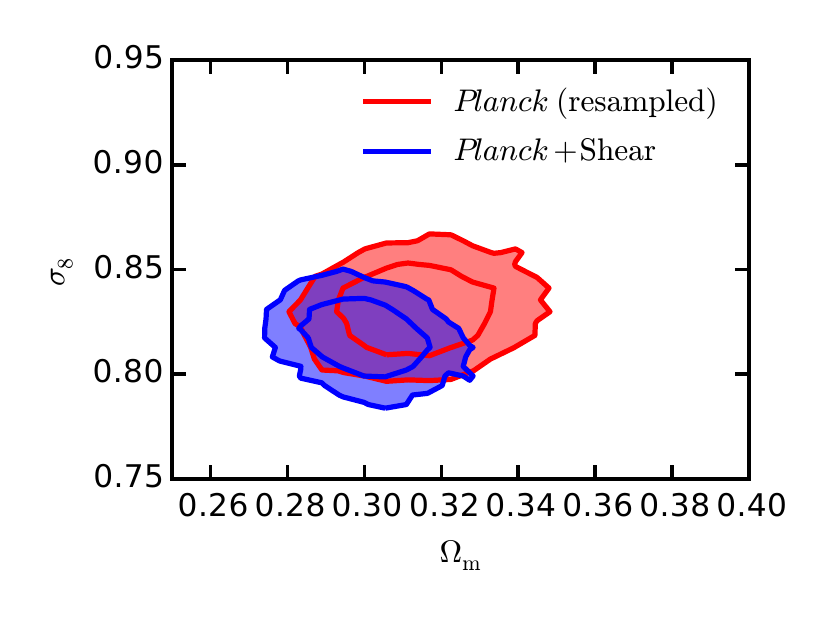}
	\caption{We show $68$ and $95$ per cent credible intervals (red, inner and outer contour, respectively) derived from sampling only the \textit{Planck} likelihood (TT+lowP) with the simplified model consisting of six cosmological parameters and only one nuisance parameter (cf. Table~\ref{tab:results2}). We combine the \textit{Planck} likelihood then with the shear likelihood and sample from the combined likelihood for the same simplified model and derive $68$ and $95$ per cent credible intervals (blue, inner and outer contour, respectively).}
	\label{fig:bl_omega_sigma_plane_planck}
\end{figure}

Prior ranges and parameter constraints for the resampled \textit{Planck} likelihood and the combination of \textit{Planck}+Shear are presented in Table~\ref{tab:results2}.
Fig.~\ref{fig:bl_omega_sigma_plane_planck} demonstrates that combining the shear likelihood with the \textit{Planck} likelihood yields improved constraints on $\sigma_8$ and $\Omega_{\rm m}$ and breaks the degeneracy between the two parameters. The 68 and 95 per cent credible intervals are largely overlapping and show marginal consistency between the two data sets as already observed above. We find the constraints $\sigma_8=0.818 \pm 0.013$ and $\Omega_{\rm m}= 0.300 \pm 0.011$ which are consistent with the constraints from the resampled \textit{Planck}-only likelihood (cf. Table~\ref{tab:results2}).

\section{Conclusions}
\label{sec:conclusions}
In this work we generalized the original quadratic estimator approach by \citet{Hu2001} to include tomographic bins. We validated the method and its extension to tomographic bins by applying it to mock data tailored to the survey specifications of CFHTLenS. In particular we made use of the official CFHTLenS Clone but produced also our own sets of Gaussian random field realizations in order to test the performance for the larger scales used in our analysis. We also used the 184 independent shear catalogues from the CFHTLenS Clone and our GRFs to derive a run-to-run covariance. We applied the method to public shear data from CFHTLenS in two tomographic bins to extract band powers of the lensing power spectrum.

We use the extracted band powers and the run-to-run covariance estimated from our suite of mock data in order to sample the shear likelihood. The sampling is performed in a Bayesian framework. We derive constraints on cosmological parameters as well as the Bayesian evidence for each model. In addition to the five baseline cosmological parameters, our most conservative model extension includes a free total mass of three degenerate massive neutrinos, a free amplitude for the baryon feedback model of the matter power spectrum and photometric redshift biases to marginalize over. For this model we derive an upper limit on the total mass of three degenerate massive neutrinos of $\Sigma m_\nu < 4.53 \, {\rm eV}$ at 95 per cent credibility. Based on the analysis of the shear likelihood we find no evidence for any of the tested model extensions though: a standard, five parameter $\Lambda$CDM model is sufficient to describe the lensing power spectrum band powers measured over a range of $80 \leq \ell \leq 1310$ in two tomographic bins. The main parameters constrained by the lensing power spectra are $\sigma_8$ and $\Omega_{\rm m}$ and we find the 68 percent credible intervals in this parameter plane to be marginally consistent both with \citet{Planck2015_CP} and the CFHTLenS analysis by \citet{Heymans2013}. Because the constraints on $\sigma_8$ and $\Omega_{\rm m}$ are degenerate, we combine both parameters into the projected parameter $S_8 \equiv \sigma_8(\Omega_\mathrm{m}/0.3)^{0.5}$. For the baseline $\Lambda$CDM model we obtain a best-fitting value of $S_8 = 0.768_{-0.039}^{+0.045}$. Marginalization over a photometric redshift bias per tomographic bin increases the errorbars on $S_8$ by $\approx 7$ per cent. Furthermore, we compare our constraints on cosmological parameters with other CFHTLenS studies and the recent result from DES and we find general agreement.
Combining the shear likelihood with the \textit{Planck} likelihood (TT+lowP) and sampling a simple six parameter $\Lambda$CDM model breaks the degeneracy between $\Omega_{\rm m}$ and $\sigma_8$ and yields the constraints $\Omega_{\rm m}=0.300 \pm 0.011$ and $\sigma_8=0.818 \pm 0.013$. These constraints are consistent with the ones derived from resampling the \textit{Planck}-only likelihood and the errorbars decrease by $\approx 19$ per cent for $\Omega_{\rm m}$ and $\approx 10$ per cent for $\sigma_8$.

Data from larger weak lensing surveys such as the Kilo-Degree Survey (KiDS\footnote{\url{kids.strw.leidenuniv.nl}}, \citealt{KiDS2013, deJong2015, Kuijken2015}), the Subaru Hyper SuprimeCam lensing survey (HSC\footnote{\url{www.naoj.org/Projects/HSC/}}), and the Dark Energy Survey (DES\footnote{\url{www.darkenergysurvey.org}}, \citealt{Flaugher2005, Jarvis2015, Becker2015}) are building up right now, and these surveys will reach full coverage in the next years. This development will culminate in the surveys carried out by the Large Synoptic Survey Telescope (LSST\footnote{\url{www.lsst.org}}, \citealt{LSST2008}) and the spaceborne \textit{Euclid}\footnote{\url{www.euclid-ec.org}}--survey \citep{Euclid}. Given these surveys, we consider our analysis also as a proof of concept in preparation for the order(s) of magnitude increase in survey area, which also implies a significant reduction in statistical uncertainties.

\section*{Acknowledgements}
The authors would like to thank the anonymous referee for her/his insightful comments that helped to further improve this work and its presentation.\\
FK would like to thank B. Audren, C. Heymans, W. Hu, J. Lesgourgues, M. Takada, and J. Torrado for fruitful discussions, technical expertise, and comments which helped to improve this work.\\
FK acknowledges support from a de Sitter Fellowship of the Netherlands Organization for Scientific Research (NWO) under grant number 022.003.013.\\
MV and HH acknowledge support from the European Research Council under FP7 grant number 279396. MV also acknowledges support from the Netherlands Organization for Scientific Research (NWO) through grants 614.001.103.\\
BJ acknowledges support by an STFC Ernest Rutherford Fellowship, grant reference ST/J004421/1.\\
The authors would like to thank the whole CFHTLenS team for their tremendous efforts and especially for their decision to make all data products available to the public. \\
Special thanks are due to C. Heymans for making her original likelihood data available to the authors.\\
Plots in this paper were prepared with {\small PYTHON}, its {\small MATPLOTLIB} package \citep{Hunter2007}, and our customized version of {\small TRIANGLE.PY} \citep{Triangle_plotter2014}.\\
% default acknowledgement as reqired by CFHTLenS data usage:
This work is based on observations obtained with MegaPrime/MegaCam, a joint project of CFHT and CEA/IRFU, at the Canada-France-Hawaii Telescope (CFHT) which is operated by the National Research Council (NRC) of Canada, the Institut National des Sciences de l'Univers of the Centre National de la Recherche Scientifique (CNRS) of France, and the University of Hawaii. This research used the facilities of the Canadian Astronomy Data Centre operated by the National Research Council of Canada with the support of the Canadian Space Agency. CFHTLenS data processing was made possible thanks to significant computing support from the NSERC Research Tools and Instruments grant program.\\
% acknowledgements for CFHTLenS mocks...
Computations for the N-body simulations were performed on the TCS supercomputer at the SciNet HPC Consortium. 
SciNet is funded by: the Canada Foundation for Innovation under the auspices of Compute Canada; 
the Government of Ontario; Ontario Research Fund - Research Excellence; and the University of Toronto.

\bibliographystyle{mn2e}
\bibliography{bibliography}

\appendix
\section{Indices and derivatives}
\label{app:indices}
In Section~\ref{sec:qe} we described the generalization of the quadratic estimator to include tomographic bins. This requires a great amount of indices in a strict notation. For brevity we switched rather quickly to a set of super-indices and we also refrained from showing the explicit forms of certain matrices. Here, we give now the explicit forms of these matrices and we also calculate the derivatives used, for example, in equations~\ref{eq:stepping} or~\ref{eq:trace} in index notation.
 
First, we start with specifying the components of the vector $\mathcal{B}$ which contains all band powers $\beta$ of type $\theta$ for each \textit{unique} redshift bin correlation $\zeta$ as $\mathcal{B}_{\zeta\theta\beta}$. Note that the total number of unique correlations between $n_{z}$ redshift bins is $n_{\rm corr} = n_{z}(n_{z}+1)/2$, because all cross-correlations contain the same information by construction.
Likewise we define the components of the tensor $\bmath{G}$ which encodes all geometric information of the field depending on the band power $\beta$, the band type $\theta$ and the redshift bin correlation $\zeta$ as
\begin{equation}
\label{eq:geometry_tensor}
\begin{split}
G_{\zeta\theta\beta (\mu\nu)(ab)(ij)} \equiv &M_{\zeta (\mu\nu)} \int_{\ell_{\rm min}(\beta)}^{\ell_{\rm max}(\beta)} \frac{\mathrm{d}\ell}{2(\ell+1)}\\
&\times \left[ w_0(\ell)I^{\theta}_{(ab)(ij)} + \tfrac{1}{2} \, w_4(\ell) Q^{\theta}_{(ab)(ij)} \right] \, . 
\end{split}
\end{equation}
We also note here that each realization of $\bmath{G}$ for a given band power $\beta$ of type $\theta$ and correlation $\zeta$ can be represented as a matrix $\bmath{{\rm G}}_{\zeta\theta\beta}$. We can write out the matrices $\bmath{{\rm I}}^{\theta}$ and $\bmath{{\rm Q}}^{\theta}$ for the EE-, BB-, and EB-band powers as \citep{Hu2001}:
\begin{align}
\bmath{{\rm I}}^{{\rm EE}} &= 
\begin{pmatrix} 
J_0+c_4J_4& s_4J_4 \\ 
s_4J_4& J_0-c_4J_4 
\end{pmatrix} \, , \label{eq:matrices_first} \\
\bmath{{\rm I}}^{{\rm BB}} &= 
\begin{pmatrix} 
J_0-c_4J_4& -s_4J_4 \\ 
-s_4J_4& J_0+c_4J_4 
\end{pmatrix} \, , \\
\bmath{{\rm I}}^{{\rm EB}} &= 
\begin{pmatrix} 
-2s_4J_4& 2c_4J_4 \\ 
2c_4J_4& 2s_4J_4 
\end{pmatrix} \, ,
\end{align}
and
\begin{align}
\bmath{{\rm Q}}^{{\rm EE}} &= 
\begin{pmatrix} 
J_0+2c_4J_4+c_8J_8& s_8J_8 \\ 
s_8J_8& -J_0+2c_4J_4-c_8J_8 
\end{pmatrix} \, , \\
\bmath{{\rm Q}}^{{\rm BB}} &= 
\begin{pmatrix} 
-J_0+2c_4J_4-c_8J_8& -s_8J_8 \\ 
-s_8J_8& J_0+2c_4J_4+c_8J_8
\end{pmatrix} \, , \label{eq:Lin_correction} \\
\bmath{{\rm Q}}^{{\rm EB}} &= 
\begin{pmatrix} 
-2s_8J_8& 2J_0+2c_8J_8 \\ 
2J_0+2c_8J_8& 2s_8J_8 
\end{pmatrix} \, . \label{eq:matrices_last}
\end{align}
In these equations we suppressed the argument of the Bessel functions $J_n$ which in each case is the product $\ell \, r_{ij}$, where $r_{ij} = |\bmath{n}_i-\bmath{n}_j|$ is the distance between pixels $i$, $j$ (cf. Section~\ref{sec:data_meas}). Moreover, we employ the shorthand notations $c_n = \cos(n\varphi)$ and $s_n = \sin(n\varphi)$, where $\varphi$ is the angle between the $x$-axis and the distance vector $\bmath{r}_{ij}$ between pixels $i$, $j$ (cf. Section~\ref{sec:data_meas}). Note that in equation~\ref{eq:Lin_correction} we corrected the misprint in the original reference pointed out by \citet{Lin2012}. Note also, that each block in the matrices of equations~\ref{eq:matrices_first} to~\ref{eq:matrices_last} defines again a matrix in the indices $i$, $j$. 

The matrices $\bmath{\rm M}_\zeta$ in equation~\ref{eq:geometry_tensor} map between the redshift bins and their unique correlations. In order to construct them, we start with the standard basis $\bmath{\rm e}_{\mu \nu}$ for $\mu \times \nu$ matrices with $\mu, \nu \in (1,...,n_z)$. For example, the standard basis for $n_z = 2$ can be written explicitly as:
\begin{align}
\bmath{{\rm e}}_{11} &= 
\begin{pmatrix}
  1 & 0 \\
  0 & 0
 \end{pmatrix} , \,
\bmath{{\rm e}}_{12} = 
\begin{pmatrix}
  0 & 1 \\
  0 & 0
 \end{pmatrix} , \, \\
\bmath{{\rm e}}_{21} &= 
\begin{pmatrix}
  0 & 0 \\
  1 & 0
 \end{pmatrix} , \, 
\bmath{{\rm e}}_{22} = 
\begin{pmatrix}
  0 & 0 \\
  0 & 1
 \end{pmatrix} . 
\end{align} 
\noindent
The index pairs $(\mu, \nu)$ can be trivially mapped to only one index $\zeta'$ which yields for the example above, i.e., $\mu, \nu \in (1, 2)$:
\begin{equation}
(1, 1) \rightarrow 1, \,  
(1, 2) \rightarrow 2, \, 
(2, 1) \rightarrow 3, \, 
(2, 2) \rightarrow 4.
\end{equation}
Imposing now, however, the symmetry condition $(\mu, \nu)=(\nu, \mu)$, which guarantees that for $n_z$ redshift bins we only consider $n_{\rm corr} = n_{z}(n_{z}+1)/2$ independent correlations, yields the symmetric mapping matrices:
\begin{align}
\bmath{{\rm M}}_{1} &= 
\begin{pmatrix}
  1 & 0 \\
  0 & 0
\end{pmatrix} = \bmath{{\rm e}}_{11} , \, \\
\bmath{{\rm M}}_{2} &= 
\begin{pmatrix}
  0 & 1 \\
  1 & 0
 \end{pmatrix} = \bmath{{\rm e}}_{12}+\bmath{{\rm e}}_{21}, \, \\
\bmath{{\rm M}}_{3} &= 
\begin{pmatrix}
  0 & 0 \\
  0 & 1
 \end{pmatrix} = \bmath{{\rm e}}_{22}\, 
\end{align} 
\noindent
which implies the following mapping from $(\mu, \nu)$ to $\zeta$:
\begin{equation}
(1, 1) \rightarrow 1, \,
(1, 2)=(2, 1) \rightarrow 2, \,
(2, 2) \rightarrow 3.
\end{equation}
\noindent
Next we construct the signal matrix $\bmath{{\rm C}}^{{\rm sig}}$ as the sum over bands $\beta$, band types $\theta$, and redshift-correlations $\zeta$ of the product of the band power vector $\bmath{\mathcal{B}}_{\zeta\theta\beta}$ with the geometry matrices $\bmath{{\rm G}}_{\beta\theta\zeta}$,
\begin{equation}
C^{{\rm sig}}_{(\mu\nu)(ab)(ij)} = \sum_{\zeta, \theta, \beta} \mathcal{B}_{\zeta\theta\beta} G_{\zeta\theta\beta(\mu\nu)(ab)(ij)} \, .
\end{equation}
\noindent
Note that the full covariance matrix $\bmath{{\rm C}}$ also includes contributions from the shape noise matrix $\bmath{{\rm C}}^{{\rm noise}}$ (cf. equation~\ref{eq:noise}) which is constant. Thus if we wish to take the derivative of the full covariance matrix with respect to every possible band power combination $B_{(\mu\nu)(\beta\theta)}$, this constant noise term vanishes and we are left with 
\begin{align}
\frac{\partial C_{(\mu\nu)(ab)(ij)}}{\partial \mathcal{B}_{\zeta\theta\beta}} &= \frac{\partial C^{\rm sig}_{(\mu\nu)(ab)(ij)}}{\partial \mathcal{B}_{\zeta\theta\beta}} \\
&= G_{\zeta\theta\beta(\mu\nu)(ab)(ij)} \nonumber \\
&\equiv \bmath{{\rm D}}_{\zeta\theta\beta} \equiv \bmath{\rm D}_A \, . \label{eq:deriv_index}
\end{align}
\noindent
In order to simplify our notation with respect to the Newton-Raphson algorithm we introduced in the last step the super-index $A$: each specific index combination $(\zeta\theta\beta)$ can be mapped to a single index $A$\footnote{For example, consider again two redshift bins, i.e., $\zeta \in (1, 2, 3)$, from which we wish to extract four band powers, i.e., $\beta \in (1, 2, 3, 4)$, of a single band type $EE \, \hat{=} \, \theta = 0$. Then we can map each unique combination of $\zeta\theta\beta$ to an integer $A \in (0, ..., 12)$.}, i.e., we denote a specific derivative matrix now as $\bmath{{\rm D}}_{A}$ instead of $\bmath{{\rm D}}_{\zeta\theta\beta}$. Hence the components of the generalized Fisher matrix $\bmath{{\rm F}}$ can be written as: 
\begin{equation}
F_{AB} = \tfrac{1}{2} \tr [\bmath{{\rm C}}^{-1} \bmath{{\rm D}}_{A} \bmath{{\rm C}}^{-1} \bmath{{\rm D}}_{B}] \, .
\end{equation}
All other equations employed in the Newton-Raphson algorithm still hold with respect to this new set of super-indices $(A, B)$.

Finally, it only remains to write out explicitly the derivatives of the full covariance matrix $\bmath{\rm C}$ with respect to the power at an integer multipole $\ell$. This is required for the calculation of the window function matrix (cf. equation~\ref{eq:window}) in which the derivatives $\bmath{\rm D}_{\ell}$ enter in computing the trace matrix $\bmath{{\rm T}}$ (cf. equation~\ref{eq:trace}):
\begin{align}
\label{eq:deriv_BWM}	 
	\frac{\partial C_{(\mu\nu)(ab)(ij)}}{\partial \mathcal{B}(\ell)} &= \sum_{\zeta, \theta}\frac{M_{\zeta(\mu\nu)}}{2(\ell+1)} [ w_0(\ell)I^{\theta}_{(ab)(ij)}\\
	& \hphantom{{} = } + \tfrac{1}{2} \, w_4(\ell) Q^{\theta}_{(ab)(ij)} ] \, \nonumber \\ 
	&\equiv D_{(\mu\nu)(ab)(ij)}(\ell) \equiv \bmath{\rm D}_{\ell} \, ,
\end{align}
where we have used that
\begin{align}
C^{{\rm sig}}_{(\mu\nu)(ab)(ij)} &= \sum_{\zeta, \theta, \ell} \mathcal{B}_{\zeta\theta}(\ell) \frac{M_{\zeta(\mu\nu)}}{2(\ell+1)} [ w_0(\ell)I^{\theta}_{(ab)(ij)}\\
& \hphantom{{} = } + \tfrac{1}{2} \, w_4(\ell) Q^{\theta}_{(ab)(ij)} ] \, \nonumber \, .
\end{align}
\section{Additional figures}
\label{app:figures}

In the following figures we show the extracted E- and B-modes for each CFHTLenS patch individually. Note that these E-mode signals enter directly in the likelihood sampling whereas the combined signal presented in Fig.~\ref{fig:signals_EE} serves just for illustrative purposes.
\begin{figure*}
	\centering
	\includegraphics[width=180mm]{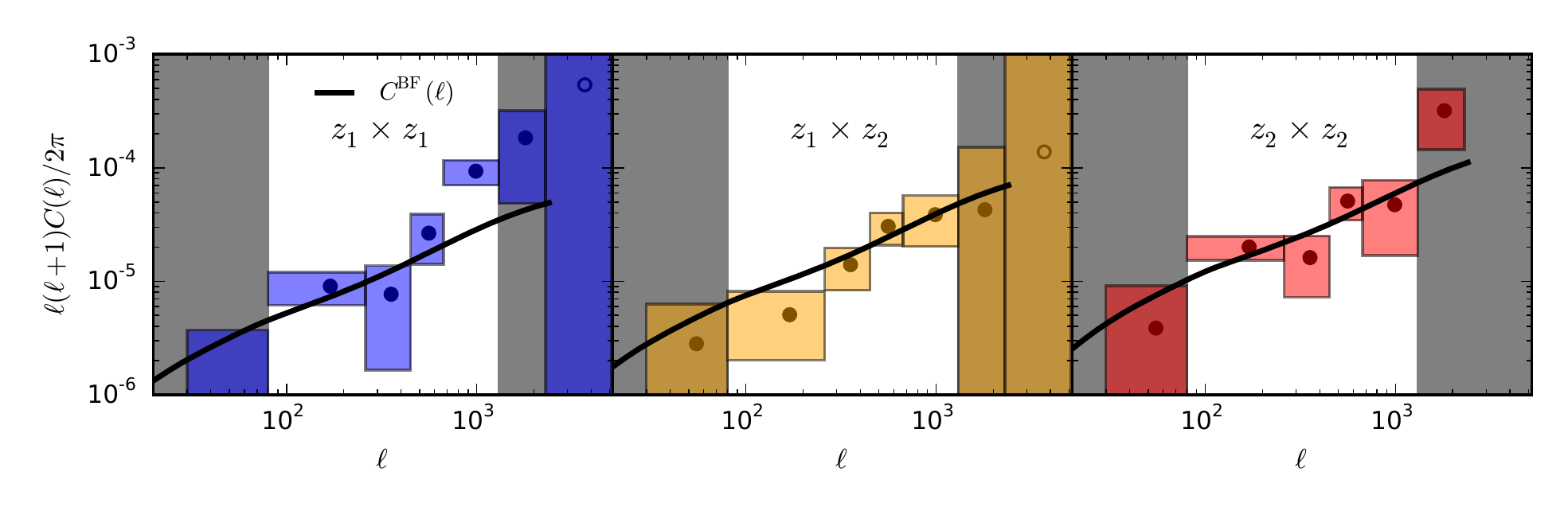}
	\caption{Measured E-mode band powers in tomographic bins for the CFHTLenS patch W1. From left to right we show the auto-correlation signal of the low redshift bin (blue), the cross-correlation signal between the low and the high redshift bin (orange), and the auto-correlation signal of the high redshift bin (red). The low redshift bin contains objects with redshifts in the range $0.5 < z_1 \leq 0.85$ and the high redshift bin covers a range $0.85<z_2\leq 1.3$. The $1\sigma$-errors in the signal are derived from a run-to-run covariance over 184 independent mock data fields (cf. Section~\ref{sec:covariance}) whereas the extend in $\ell$-direction is the width of the band. Band powers in the shaded regions (grey) to the left and right of each panel are excluded from the cosmological analysis (cf. Fig.~\ref{fig:signals_BB}). The solid line (black) shows the power spectrum for the best fitting, five parameter $\Lambda$CDM model derived in the subsequent analysis (cf. Section~\ref{sec:cosmo_inference} and Table~\ref{tab:results1}). Note, however, that the band powers are centred at the naive $\ell$-bin centre and thus the convolution with the band window function is not taken into account in this plot, in contrast to the cosmological analysis.} 
	\label{fig:signals_EE_W1}
\end{figure*}
\begin{figure*}
	\centering
	\includegraphics[width=180mm]{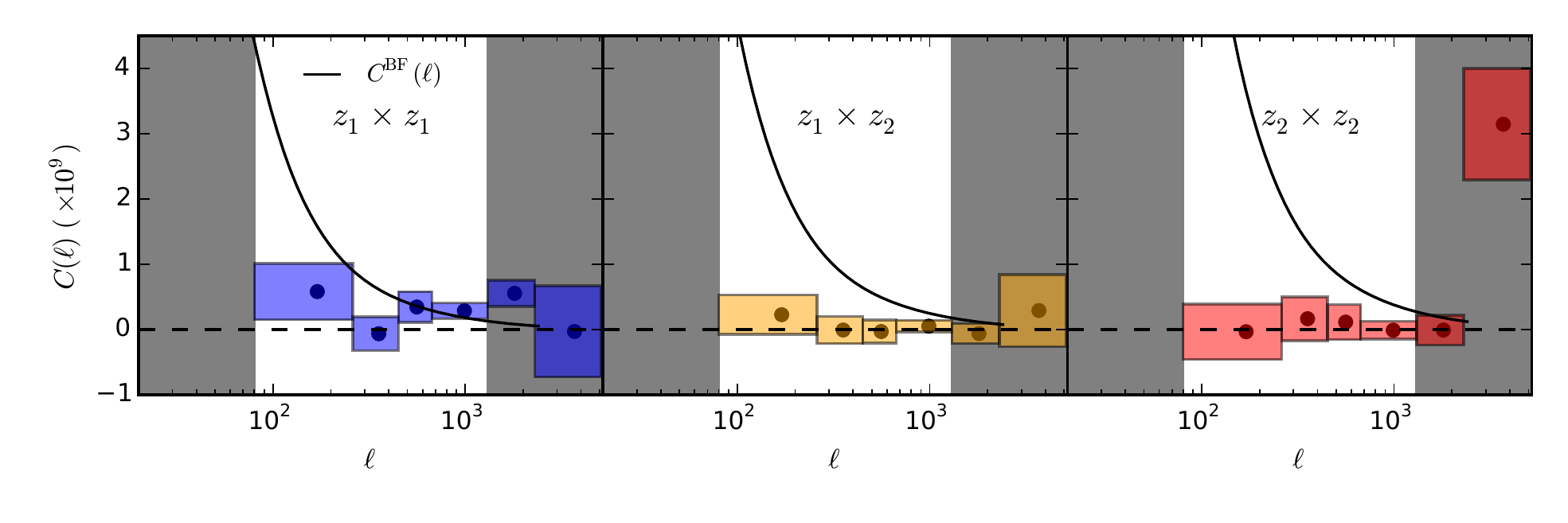}
	\caption{Same as Fig.~\ref{fig:signals_EE_W1} but for B-mode band powers. Note, however, the different scale (linear) and normalization used here with respect to Fig.~\ref{fig:signals_EE_W1}; for reference we also plot the best fitting E-mode power spectrum as solid line (black). We show the measured B-modes as (black) dots with $1\sigma$-errors derived from the inverse Fisher matrix. Based on these signals we define the shaded regions (grey) to the left and right of each panel (cf. Section~\ref{sec:signal} for details). E-mode band powers in these regions are excluded from the cosmological analysis (cf. Fig.~\ref{fig:signals_EE}).}
	\label{fig:signals_BB_W1}
\end{figure*}
\begin{figure*}
	    \centering
		\includegraphics[width=180mm]{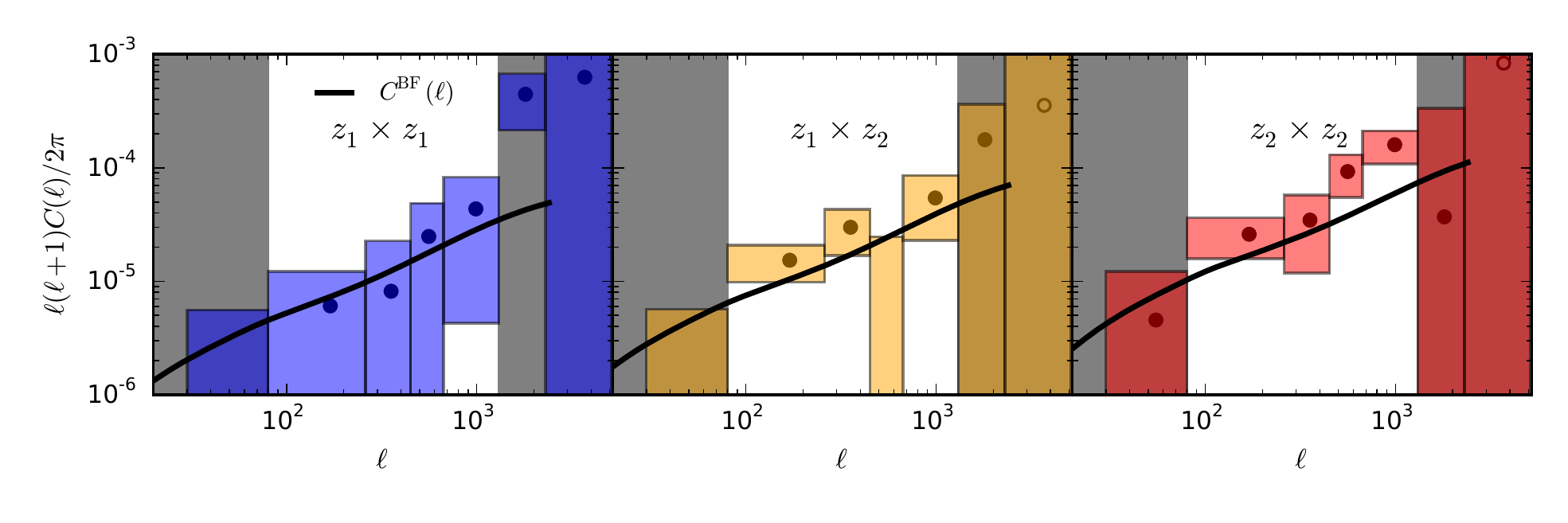}
		\label{fig:signals_EE_W2}
		\caption{Same as Fig.~\ref{fig:signals_EE_W1} but for CFHTLenS patch W2. Open symbols denote negative values plotted with $1\sigma$-errors centred on the absolute value.}
\end{figure*}
\begin{figure*}
	    \centering
		\includegraphics[width=180mm]{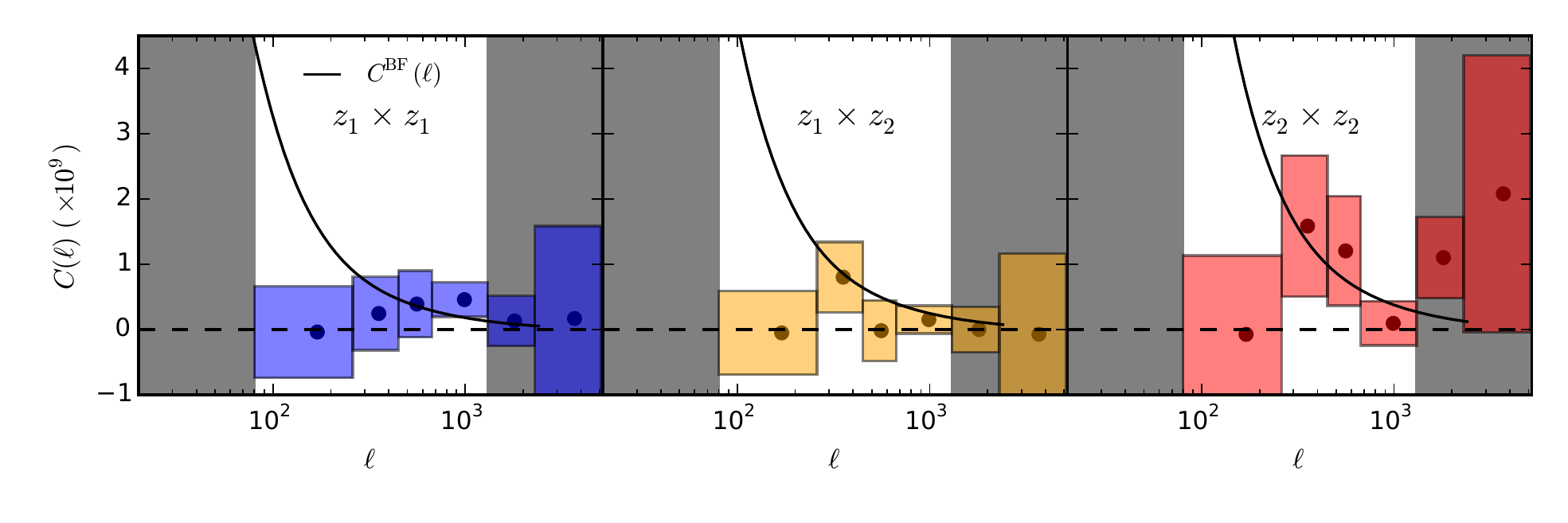}
		\label{fig:signals_BB_W2}
		\caption{Same as Fig.~\ref{fig:signals_BB_W1} but for CFHTLenS patch W2. Open symbols denote negative values plotted with $1\sigma$-errors centred on the absolute value.}
\end{figure*}
\begin{figure*}
	    \centering
		\includegraphics[width=180mm]{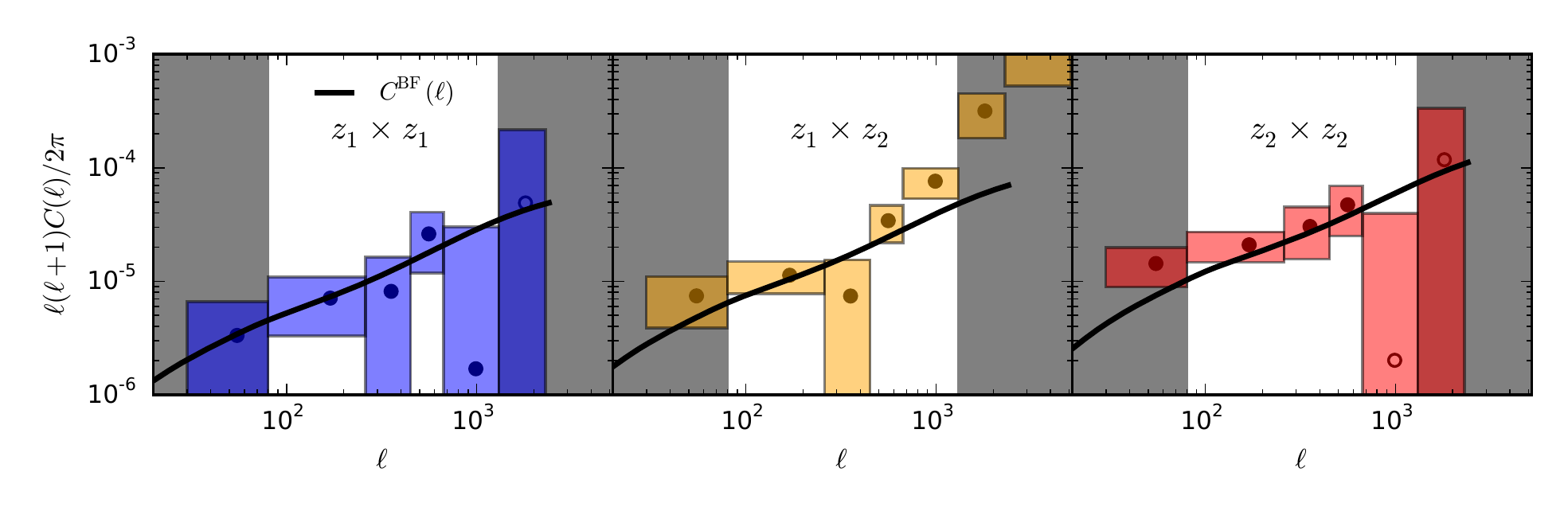}
		\label{fig:signals_EE_W3}
		\caption{Same as Fig.~\ref{fig:signals_EE_W1} but for CFHTLenS patch W3. Open symbols denote negative values plotted with $1\sigma$-errors centred on the absolute value.}
\end{figure*}
\begin{figure*}
	    \centering
		\includegraphics[width=180mm]{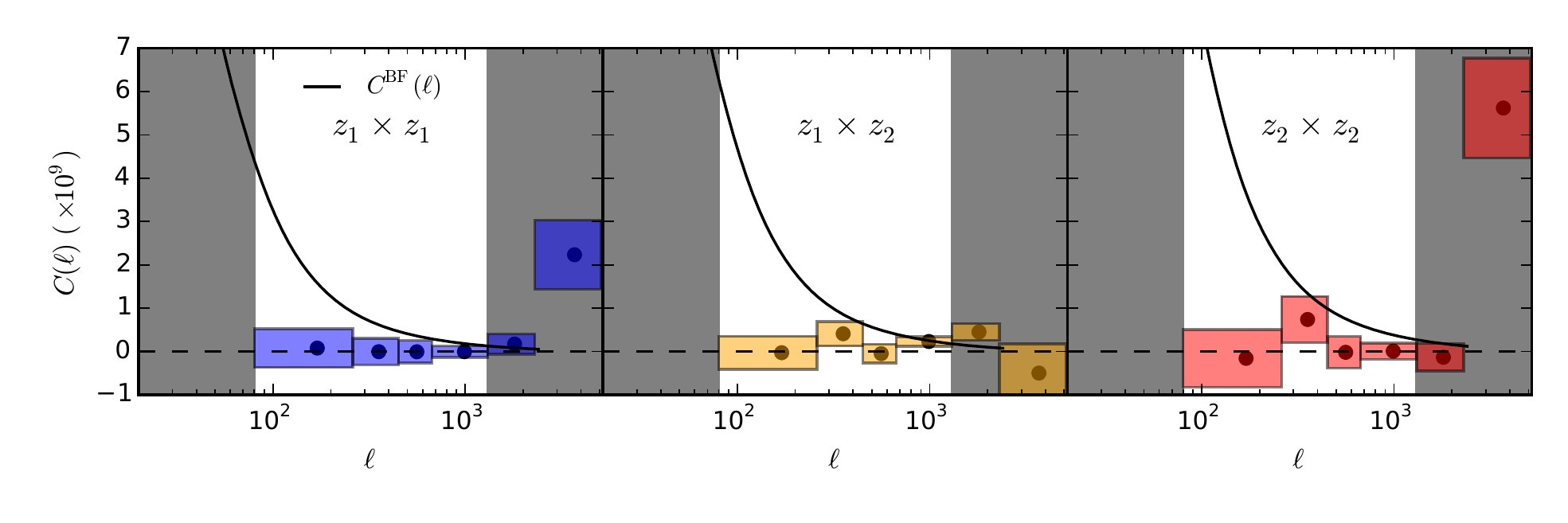}
		\label{fig:signals_BB_W3}
		\caption{Same as Fig.~\ref{fig:signals_BB_W1} but for CFHTLenS patch W3. Open symbols denote negative values plotted with $1\sigma$-errors centred on the absolute value.}
\end{figure*}
\begin{figure*}
	    \centering
		\includegraphics[width=180mm]{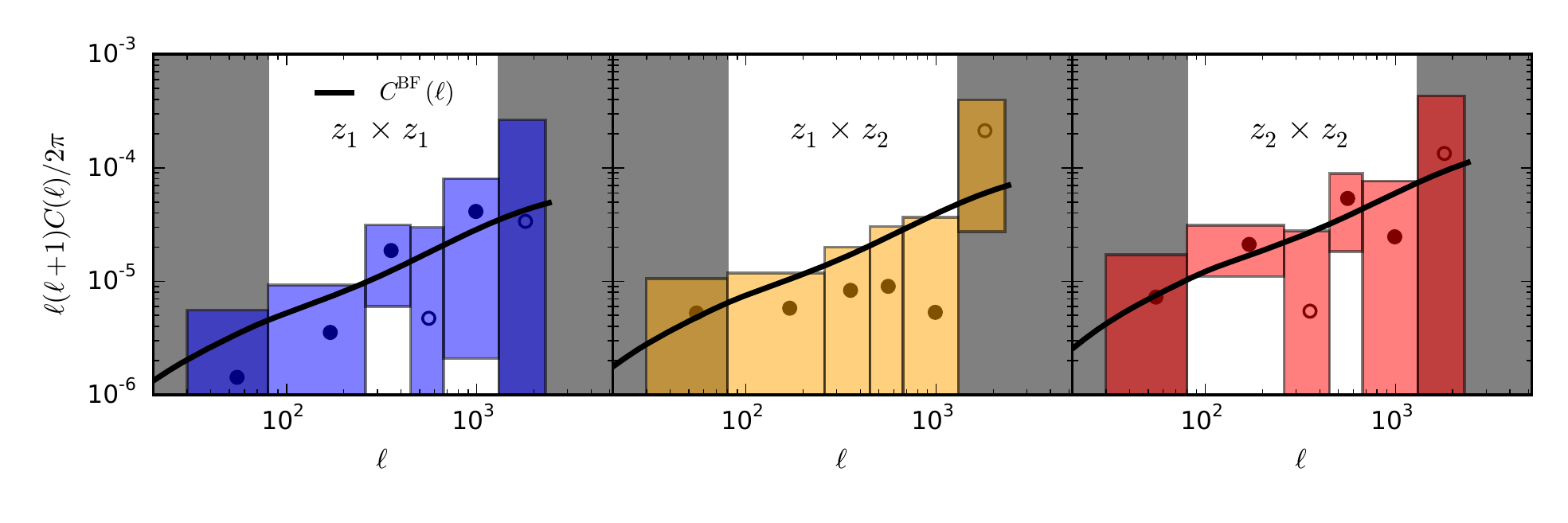}
		\label{fig:signals_EE_W4}
		\caption{Same as Fig.~\ref{fig:signals_EE_W1} but for CFHTLenS patch W4. Open symbols denote negative values plotted with $1\sigma$-errors centred on the absolute value.}
\end{figure*}
\begin{figure*}
	    \centering
		\includegraphics[width=180mm]{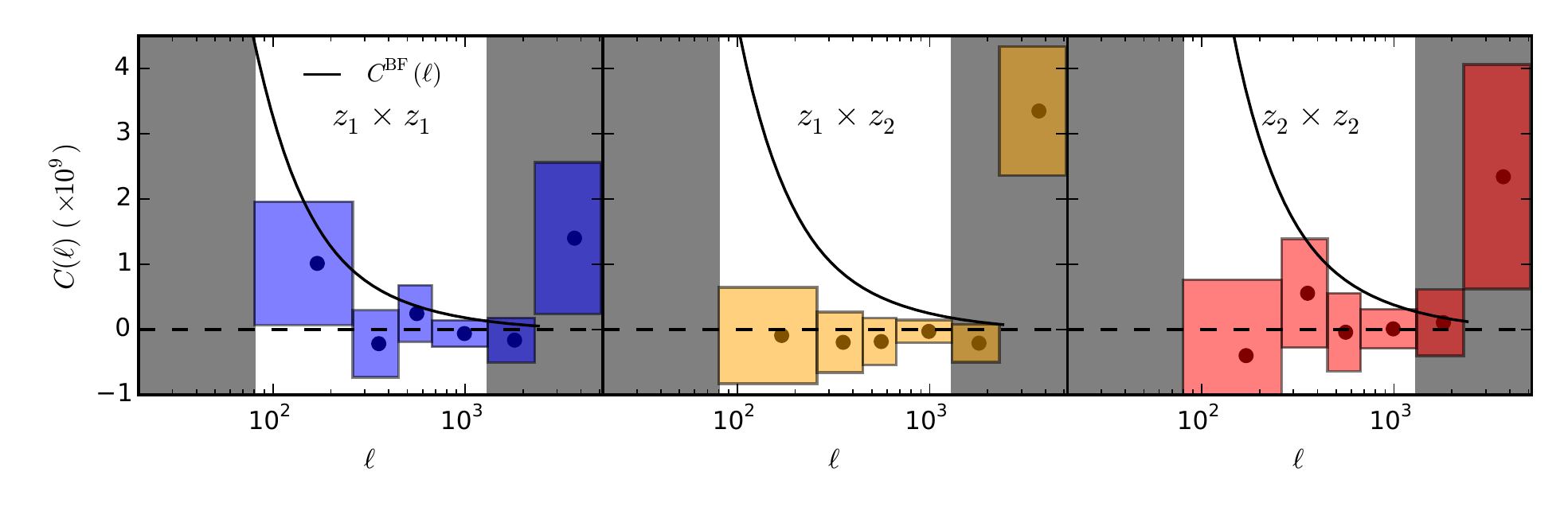}
		\label{fig:signals_BB_W4}
		\caption{Same as Fig.~\ref{fig:signals_BB_W1} but for CFHTLenS patch W4. Open symbols denote negative values plotted with $1\sigma$-errors centred on the absolute value.}
\end{figure*}

\label{lastpage}
\end{document}